\documentclass[paper]{JHEP3} % 10pt is ignored!

\JHEPspecialurl{http://jhep.sissa.it/JOURNAL/JHEP3.tar.gz}

\usepackage{epsfig,multicol}
\usepackage{cite}
\usepackage{amsmath}
\usepackage{amssymb}
\preprint{
KEK-TH-971\\
IU-TH-1 \\
hep-th/0506062\\}

\title{Impact of supersymmetry on the 
nonperturbative dynamics of fuzzy spheres}
\author{ Konstantinos~N.~Anagnostopoulos$^a$, Takehiro~Azuma$^b$, 
Keiichi~Nagao$^{b,c}$ and Jun~Nishimura$^{b,d}$\\
\llap{$^a$}Physics Department, National Technical University, \\
Zografou Campus, GR-15780 Athens, Greece \\
\llap{$^b$}High Energy Accelerator Research Organization (KEK),\\
1-1 Oho, Tsukuba 305-0801, Japan  \\
\llap{$^c$}Theoretical Physics Laboratory, College of Education, \\
Ibaraki University, 2-1-1 Bunkyo, Mito, Ibaraki 310-8512, Japan \\
\llap{$^d$}Department of Particle and Nuclear Physics,\\
Graduate University for Advanced Studies (SOKENDAI),\\
1-1 Oho, Tsukuba 305-0801, Japan  \\
\email{konstant@physics.uoc.gr,
azumat@post.kek.jp, 
nagao@mx.ibaraki.ac.jp, 
jnishi@post.kek.jp}} 

\abstract{
We study 
a 4d supersymmetric matrix model
with a cubic term, which incorporates fuzzy spheres 
as classical solutions, using Monte Carlo simulations
and perturbative calculations.
The fuzzy sphere in the supersymmetric model
turns out to be always stable
if the large-$N$ limit is taken in such a way that
various correlation functions scale.
This is
in striking contrast to analogous bosonic models,
where the fuzzy sphere decays into the pure Yang-Mills vacuum
due to quantum effects when the coefficient of the cubic term 
becomes smaller than a critical value.
We also find that the power-law tail of the eigenvalue 
distribution, which exists in the supersymmetric model 
{\em without} the cubic term,
disappears in the presence of the fuzzy sphere in the large-$N$ limit.
Coincident fuzzy spheres turn out to be unstable, which implies
that the dynamically generated gauge group is U(1).
}


\newcommand{\bel}{\begin{equation}\label}

\newcommand{\n}{\nonumber}
\newcommand {\beq}{\begin{equation}}
\newcommand {\eeq}{\end{equation}}
\newcommand {\beqa}{\begin{eqnarray}}
\newcommand {\eeqa}{\end{eqnarray}}
\newcommand {\bc}{\begin{center}}
\newcommand {\ec}{\end{center}}
\newcommand {\tr}{{\rm tr\,}}
\newcommand {\Tr}{\mbox{Tr\,}}

\newcommand {\Det}{\mbox{Det}}

\newcommand {\ee}{\mbox{e}}

\newcommand {\dd}{\mbox{d}}

\newcommand {\stwo}{{\rm S}^{2}}

\def\dag{\dagger}

\def\vs5{\vspace*{5mm}}
\def\vs1{\vspace*{1cm}}
\def\vs2{\vspace*{2cm}}
\def\hs5{\vspace*{5mm}}
\def\hs1{\hspace*{1cm}}
\def\hs2{\hspace*{2cm}}
\def\vs50{\vspace*{50mm}}
\def\vs20{\vspace*{20mm}}

\def\tr{\hbox{tr}}

\begin{document}

\section{Introduction}

Matrix models are considered as one of the most promising candidates
for a nonperturbative formulation of string theories.
%, and it is expected to play the role of lattice gauge theories 
%in field theories.
Indeed some concrete models are proposed
as constructive definitions of 
superstring and M theories \cite{9610043,9612115}.
%All of t
These models are obtained from the 
dimensional reduction of super Yang-Mills theory in ten dimensions.
In a broad sense, such models belong to the class of the 
so-called dimensionally reduced models \cite{Eguchi:1982nm}
(or large-$N$ reduced models),
which were studied intensively in the eighties 
as an equivalent description of large-$N$ gauge theories.
Unlike the old models, however,
the new models are written in terms of Hermitian matrices,
and they have manifest supersymmetry, which is expected to have crucial
effects on their dynamics
\footnote{See ref.~\cite{Nishimura:2003rj}
for a comprehensive review on these issues.}.
%It is important to understand the dynamical properties of the 
%supersymmetric large-$N$ reduced models, since it is expect that
%we may understand the properties of the true vacuum in superstring theory.
%
%such as the space-time dimensionality, the gauge group and the 
%matter contents.

An important feature of these matrix models 
is that the space-time is not introduced 
from the outset,
%in the conventional way, 
but it emerges dynamically as the eigenvalue distribution of the
bosonic matrices. 
In fact there are certain evidences in the IIB matrix model \cite{9612115}
that {\em four-dimensional} space-time is generated dynamically 
  \cite{Nishimura:2001sx,KKKMS,Kawai:2002ub,0307007}.
In refs.~\cite{Nishimura:2001sx,KKKMS,Kawai:2002ub} 
the free energy of space-time
with various dimensionality has been calculated
using the gaussian expansion method, and the free energy turned out to
take the minimum value for the four-dimensional space-time.
In ref.~\cite{0307007} it was found that
the fuzzy ${\rm S}^2 \times {\rm S}^2$ (but not the fuzzy S$^{2}$)
is a solution to the 2-loop effective action.
See also refs.\ \cite{Aoki:1998vn,9811220,Ambjorn:2000bf,%
Ambjorn:2000dx,NV,Burda:2000mn,%
Ambjorn:2001xs,exact,sign,Vernizzi:2002mu,%
Nishimura:2004ts,Bal:2004ai}
for related works on this issue.
% See also
% refs.~\cite{Ambjorn:2000dx}--\cite{Vernizzi:2002mu} for related works.
%
%The dynamical generation of the 4d space-time requires the 
%SO($10$) symmetry to be spontaneously broken,
%where supersymmetry is expected to play a crucial role.
%
%In fact in the bosonic models it is known that the SO($D$) symmetry
%is not spontaneously broken.

By adding a Chern-Simons term to the matrix models,
one obtains fuzzy spheres \cite{Madore}
as classical solutions \cite{Alekseev:2000fd,0101102,0204256,0207115,0301055},
and their dynamical properties have been studied
in refs.\ \cite{0108002,0205213,Valtancoli:2002rx,0303120,Steinacker:2003sd,%
0309082,0312241,0402044,0403242,0412052,0412312,Kaneko:2005pw,badis}.
This provides a matrix description of the 
so-called Myers effect in string theory \cite{Myers:1999ps}.
The emergence of a fuzzy sphere in matrix models
may be regarded as a prototype of the dynamical generation of
space-time since it has lower dimensionality
than the original dimensionality that the model can actually describe.
When $k$ fuzzy spheres coincide, the gauge symmetry enhances
from U(1)$^k$ to U($k$).
By expanding the theory around such a solution, 
one obtains a U($k$) gauge theory on a noncommutative 
geometry \cite{0101102}.
Therefore the model may also serve as a toy model for
the dynamical generation of gauge group, which is expected to occur
in the IIB matrix model \cite{9903217}.

In fact one can use the above matrix models 
to define a regularized field theory on the fuzzy sphere 
as has been done on a noncommutative 
torus \cite{AMNS}, which enables nonperturbative studies of such
theories from first principles \cite{simNC}. 
%Studying field theories on a noncommutative geometry
This is motivated from the general expectation 
that noncommutative geometry provides
a crucial link to string theory \cite{Seiberg:1999vs}
and quantum gravity \cite{gravity}.
Yet another motivation 
%for studying fuzzy spheres,
is to use the fuzzy sphere (or its generalization
\cite{Dolan:2003kq,O'Connor:2003aj})
%one may use them
as a regularization scheme alternative to 
the lattice regularization \cite{Grosse:1995ar}.
Unlike the lattice, fuzzy spheres preserve the continuous symmetries 
of the space-time considered, and hence it is expected to
ameliorate the well-known problem
concerning chiral symmetry 
\cite{Grosse:1994ed,%
Carow-Watamura:1996wg, chiral_anomaly, non_chi,balagovi,% 
chiral_anomaly2,balaGW,Nishimura:2001dq,AIN,AIN2,Ydri:2002nt,% 
Iso:2002jc,Balachandran:2003ay,nagaolat03, AIN3,0412052} 
and supersymmetry.
% in lattice field theories may become easier to overcome.
% In both contexts, it is important to study four-dimensional
% fuzzy manifolds eventually in order to 
% Although the simplest fuzzy sphere is the fuzzy 2-sphere,
A challenge in this direction is to remove the effects of 
noncommutativity of the space-time in the ``continuum limit''.
The fuzzy sphere is also useful in the
Coset Space Dimensional Reduction
\cite{Forgacs:1979zs,Kapetanakis:1992hf},
where one can take the compact part of space-time to be a fuzzy coset
\cite{Aschieri:2003vy,Aschieri:2004vh}.
% and the CSDR constraints have been solved for the fuzzy
% sphere. 
%
%% It is interesting that the CS term appears in these models in
%% the potential of the coset space degrees of freedom in the four
%% dimensional effective action.

Whatever the motivation is, the stability of fuzzy-sphere-like solutions
is clearly one of the most important issues. In cases when
there are more than one stable solutions, one can
identify the true vacuum by comparing the corresponding free energy.
This will be important in the dynamical determination
of the space-time dimensionality and the gauge group
in superstring theory.
In the series of papers 
\cite{0401038,0405096,0405277,0410263,0504217,0506205},
we addressed such issues in various kinds of models
using both perturbative calculations and 
%a fully nonperturbative method based on 
Monte Carlo simulations.
In ref.\ \cite{0401038} we have studied
the dimensionally reduced 3d Yang-Mills models with the
Chern-Simons term, which has the fuzzy 2-sphere (S$^{2}$)
as a classical solution \cite{0101102}.
We have found a first-order phase transition as we vary 
the coefficient ($\alpha$) of the Chern-Simons term.
For small $\alpha$
the large-$N$ behavior of the model is the same as in the 
pure Yang-Mills model, whereas for large $\alpha$
a single fuzzy S$^2$ appears dynamically.
%The two phases are separated by a first order phase transition.
In addition we find that 
the $k$ coincident fuzzy spheres, which are also classical solutions
of the same model, cannot be realized as the true vacuum
in this model even in the large-$N$ limit.
This implies that the dynamical gauge group is U(1) in this model.
In refs.~\cite{0405096,0405277,0506205}
we have extended this work to various matrix
models, which incorporate four-dimensional fuzzy manifolds
%(i.e., fuzzy S$^4$ , CP$^2$ \cite{0405277}
%and $\mbox{S}^2 \times \mbox{S}^2$ \cite{0506205})
as classical solutions.
While the fuzzy S$^4$ turned out to be unstable \cite{0405096},
we find that the fuzzy CP$^2$ \cite{0405277} and 
the fuzzy $\mbox{S}^2 \times \mbox{S}^2$ \cite{0506205}
are stable at large $N$ although the true vacuum is actually
given by the fuzzy S$^2$.
In the latter two cases the gauge group generated dynamically turned
out to be U(1) as well. 
In ref. \cite{0504217}, on the other hand, it has been shown for the first
time that gauge groups of higher rank can be realized in the true
vacuum by adding a mass term to the 3d Yang-Mills-Chern-Simons model.

The aim of the present paper is to study the impact of supersymmetry 
on the dynamics of the fuzzy spheres.
The simplest 3d supersymmetric model \cite{0101102}
is problematic nonperturbatively since the partition function is divergent
\cite{9803117,9804199,Austing:2001bd,Austing:2001pk,0310170}.
This leads us to study the 4d supersymmetric model
with a cubic term instead.
Indeed it turns out that the supersymmetry has a striking impact.
Unlike the bosonic models,
the fuzzy sphere is always stable 
if the large-$N$ limit is taken in such a way that
various correlation functions scale.
%keeping the coefficient of the 
%Chern-Simons term fixed with an appropriate normalization constant.
We also observe an interesting phenomenon that
the power-law tail of the eigenvalue 
distribution, which exists in the supersymmetric models
{\em without} the Chern-Simons term
\cite{9902113,Austing:2001pk},
disappears in the presence of the fuzzy sphere
in the large-$N$ limit.
% where the use of an ``exact'' simulation based
%on the Hybrid Monte Carlo algorithm is crucial.
Coincident fuzzy spheres turn out to be unstable, which implies
that the dynamically generated gauge group is U(1) 
in the present model.

This paper is organized as follows.
In section \ref{section:model} we define the model and discuss
its fuzzy sphere solutions.
In section \ref{section:phase} we study the phase diagram of the model.
In section \ref{a0} we study the geometrical structure of the dominant
configurations.
In section \ref{section:multi} we study 
%the instability of 
coincident fuzzy spheres and discuss the dynamical gauge group.
Section \ref{section:summary} is devoted to summary and discussions.
In appendix \ref{algorithm} we explain the algorithm for our Monte Carlo 
simulations. In appendices \ref{one-loop-eff-action} and
\ref{one-loop-analysis} we provide the details 
of perturbative calculations.
 
\section{The model and the fuzzy sphere} 
\label{section:model}
The model we study in this paper is defined by the action
  \begin{eqnarray}
 \label{total_action}
   S &=& S_{\rm b} + S_{\rm f} \ , \\
   S_{\rm b} &=& N \, \tr \, \left( - \frac{1}{4} \, 
   [ \, A_{\mu}, \, A_{\nu} \, ]^{2}
   + \frac{2}{3} \, i \, \alpha 
   \sum_{i,j, k=1}^{3} \, \epsilon_{ijk} \, 
     A_{i} \, A_{j} \, A_{k} \right), \label{boson-definition} \\
   S_{\rm f} &=& -N  \, \tr \, \Bigl( \, {\bar \psi}_{\alpha} \, 
  (\Gamma_{\mu})_{\alpha \beta} \, 
     [ \, A_{\mu}, \, \psi_{\beta} \, ] \, \Bigr) \ , 
  \label{fermion-definition}
  \end{eqnarray}
 where $A_{\mu}$ ($\mu=1,2,3,4$) are $N \times N$ traceless hermitian 
 (bosonic) matrices, 
 and $\psi_{\alpha}, {\bar \psi}_{\alpha}$ ($\alpha = 1,2$)
 are $N \times N$ traceless complex (fermionic) matrices. 
% The spinor indices $\alpha, \beta$ run over $\alpha, \beta = 1,2$.
 Here and henceforth we assume that
 repeated Greek indices are summed over all possible integers.
 The $\epsilon_{ijk}$ ($i,j,k=1,2,3$) is a totally anti-symmetric tensor 
 with $\epsilon_{123} = 1$.
 The $2 \times 2$ matrices $\Gamma_{\mu}$ are Weyl-projected gamma matrices
 in four dimensions, and they are given explicitly as
% $\Gamma _ i = \sigma_i$, $\Gamma _4 = i \, {\bf 1}$ in
% terms of Pauli matrices $\sigma_i$. 
   \begin{eqnarray}
    \Gamma_{1} = 
% \sigma_{1} = 
\left( \begin{array}{cc} 0 & 1 \\ 1 & 0 
    \end{array} \right), \hspace{2mm}
    \Gamma_{2} =  
%\sigma_{2} = 
\left( \begin{array}{cc} 0 & -i \\ i & 0
    \end{array} \right), \hspace{2mm}
    \Gamma_{3} =  
%\sigma_{3} = 
\left( \begin{array}{cc} 1 & 0 \\ 0 & -1
    \end{array} \right), \hspace{2mm}
    \Gamma_{4} =  
%i {\bf 1} = 
\left( \begin{array}{cc} i & 0 \\ 0 & i
    \end{array} \right) \ . 
\label{gamma-matrices}
   \end{eqnarray}

 The convergence of the integration over $A_\mu$ is a non-trivial issue
 since the integration region is non-compact.
 At $\alpha=0$ the partition function is finite for arbitrary
 $N$, as first conjectured by ref.\ \cite{9803117} and 
 proved later by ref.\ \cite{Austing:2001pk}, 
%\footnote{See refs.\ \cite{9804199,Austing:2001bd} 
% for the corresponding statement in the bosonic case.}
 and this remains to be the case also 
 for $\alpha \neq 0$ \cite{0310170}.
 Moreover, since the fermion determinant of this model is positive 
 semi-definite (See ref.\ \cite{Ambjorn:2000bf} for a proof),
 the model can be studied by Monte Carlo simulations without confronting
 the so-called sign problem.
 The pure super Yang-Mills model ($\alpha = 0$), which may be regarded as
 % $\alpha = 0$ case
 % , pure super Yang-Mills model, which may be regarded as 
 the 4d version of the IIB matrix model \cite{9612115}, 
 has been studied intensively
 \cite{9803117,Ambjorn:2000bf,Burda:2000mn,Ambjorn:2001xs}.
 The sign problem does not occur even if
 one includes the cubic term, which is real 
 \footnote{This is in contrast to the Chern-Simon term in ordinary 
gauge theories in euclidean space-time, which is purely imaginary.
Note that the coefficient $\alpha$ in (\ref{boson-definition})
should be chosen to be real in order for fuzzy sphere solutions to exist.}.

%Let us discuss the symmetries of this model.
%% First of all it has the SU($N$) invariance
%%   \begin{eqnarray}
%%    A_{\mu} \to V \, A_{\mu} \, V^{\dagger}, \hspace{2mm}
%%    \psi \to V \, \psi \, V^{\dagger}, \hspace{2mm}
%%    {\bar \psi} \to V \, {\bar \psi} \, V^{\dagger} \ , 
%%   \label{suntr}
%%   \end{eqnarray}
%%   where $V \in {\rm SU}(N)$. 
  For $\alpha = 0$ the model has manifest
 SO(4) symmetry and supersymmetry.
%%   \begin{eqnarray}
%%  & &  \delta^{(1)} A_{\mu} = i {\bar \epsilon}_{1} \Gamma_{\mu} \psi,
%%   \hspace{2mm}
%%       \delta^{(1)} \psi = \frac{i}{2} \left( [A_{\mu}, A_{\nu}]
%% %   - i \alpha \epsilon_{\mu \nu \rho} A_{\rho} 
%% \right) 
%%    \Gamma_{\mu \nu} \epsilon_{1} \ , \label{hom-susy} \\
%%  & & \delta^{(2)} A_{\mu} = 0 \ , \hspace{2mm}
%%      \delta^{(2)} \psi = \epsilon_{2} \ .  
%%   \end{eqnarray}
The cubic term in (\ref{boson-definition})
obviously breaks the SO(4) symmetry down to SO(3).
It also breaks supersymmetry, but the effects of breaking is ``soft''
since the power of $A_\mu$ is lower than the quartic term \cite{0303120}.
Therefore one may still anticipate to see peculiar effects of
supersymmetry.
We repeat that the 3d supersymmetric model,
which has been studied perturbatively \cite{0101102,0303120},
is actually problematic nonperturbatively
since the partition function is divergent
\cite{9803117,Austing:2001pk,0310170}.
Therefore, the present 4d model is the simplest
model that can be studied in order to examine
the impact of supersymmetry on the fuzzy sphere dynamics.

Let us then consider the classical solutions of this model.
For $\psi = 0$ the equation of motion reads
  \begin{eqnarray}
% \left\{ \begin{array}{ll}
&~&  [ \,  A_{\nu}, [ \,  A_{\nu},  A_{i} \,  ]  \, ] 
 + \, i \, \alpha \sum_{j,k=1}^{3} \epsilon_{ijk} \,
  [  \, A_{j} ,  A_{k} \,  ] 
 =  0  \mbox{~~~for~} i=1,2,3 \ , \nonumber \\
%  \end{eqnarray}
%  \begin{eqnarray}
&~&  [ \, A_\nu , [ \, A_\nu , A_4 \, ] \,  ]  = 0 \ .
%\end{array} 
%\right. 
\label{eom}
 \end{eqnarray}
Apart from the solution given by commuting matrices,
which exists also for $\alpha = 0$,
we have the fuzzy $\stwo$ solution given by
  \begin{eqnarray}
   \left\{ \begin{array}{ll} 
      A^{(\stwo)}_{i} = \alpha \, L_{i}^{(N)} &
  \textrm{~~~for $i=1,2,3$}  \ , \\ 
      A^{(\stwo)}_{4} = 0 \ ,  &  
  \end{array} \right. \label{fs-solution}
  \end{eqnarray}
where $L_{i}^{(r)}$ ($i = 1,2,3$) 
represents the $r$-dimensional {\em irreducible} 
representation of the SU$(2)$ Lie algebra
\beq
   [L_{i}^{(r)} , L_{j}^{(r)} ] = 
   i \, \alpha \, \epsilon_{ijk} \, L_{k}^{(r)} \ .
\eeq
The solution $A^{(\stwo)}_{\mu}$ satisfies
  \begin{eqnarray}
   \sum_{i=1}^{3} \, \Bigl( A^{(\stwo)}_{i} \Bigr) ^{2} = 
  \frac{1}{4} \, (N^{2}-1) \,  \alpha^{2} \, 
  {\bf 1}_{N} \ ,
  \label{q3-casimir}
  \end{eqnarray}
 which implies that the ``radius'' of the fuzzy sphere is given by 
\beq
\rho =  \frac{1}{2} \, \alpha \, \sqrt{N^2 - 1} \ .
\eeq
We consider more general solutions in section \ref{section:multi}.

%% Thus the configuration has the geometry of the sphere, when we
%% interpret 
%% Since $A_\mu$ are not commutative to each other, the configuration
%% is called the fuzzy sphere.

\section{Phase diagram}
\label{section:phase}

\subsection{Monte Carlo results}
\label{section:MCresults}

In this section we calculate various quantities by Monte Carlo simulation
and study the phase diagram of the model (\ref{total_action}).
%first order phase transition, which occurred 
%in the bosonic models, does not appear in the present supersymmetric case.
We show results obtained by using the fuzzy sphere $A^{(\stwo)}_{\mu}$ 
as the initial configuration, but we have checked that
the result is the same for other initial configurations
such as $A_\mu=0$ or some randomly generated configurations.
For brevity we introduce the notation
\begin{eqnarray}
F_{\mu\nu} &=& i \, [A_{\mu}, A_{\nu}] \ , \\
M &=& \frac{2}{3\, N} \, i \, \sum_{i,j,k=1}^{3} 
\, \epsilon_{ijk} \, \tr \,  ( \, A_{i} \, A_{j} \, A_{k} \, ) \ .
\end{eqnarray}
We note that there is an exact result
\begin{eqnarray}
\left\langle \frac{1}{N} \, \tr \, (F_{\mu\nu})^2 \right\rangle
+ 3 \, \alpha \,  \langle M \rangle =
6 \left(1 -\frac{1}{N^2} \right) \  ,
\label{SD2}
\end{eqnarray}
which can be derived as in the bosonic case \cite{0401038}.
This result has been used to check our code for the simulation.

    \FIGURE{
    \epsfig{file=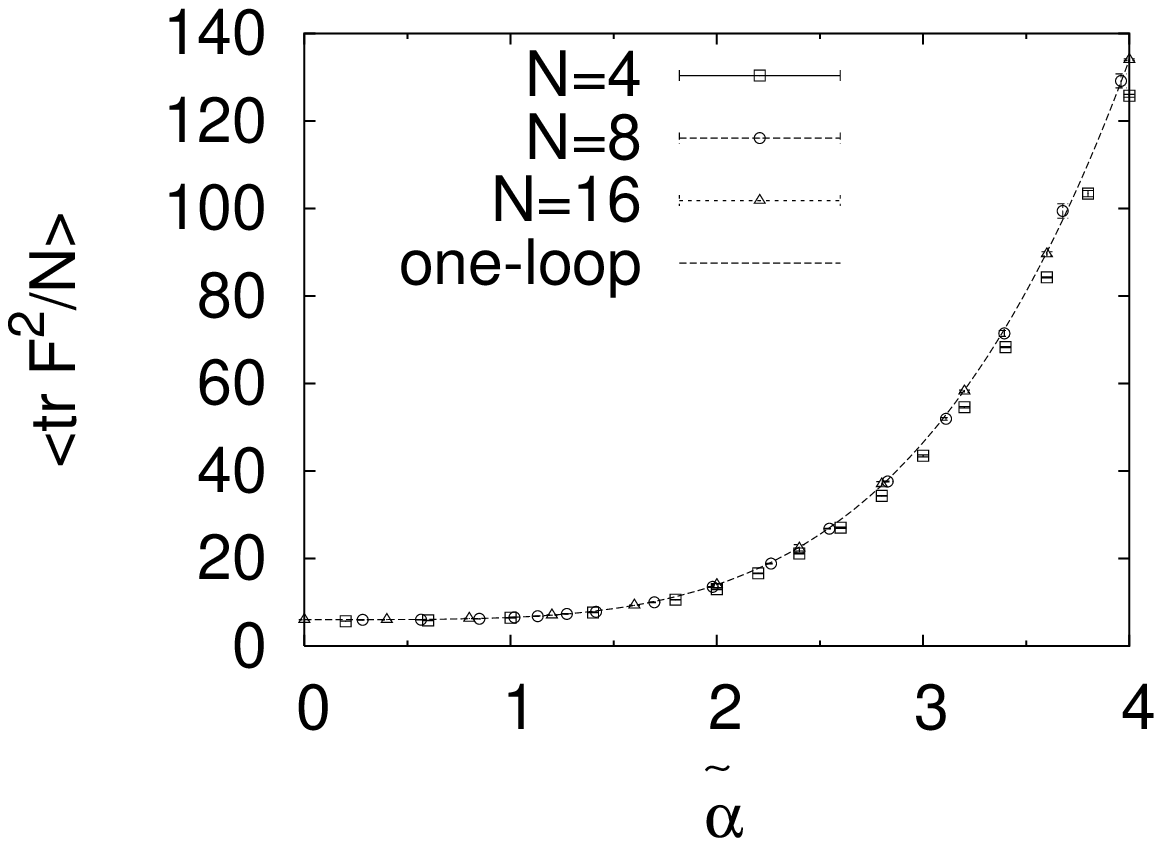,width=7.4cm}
    \epsfig{file=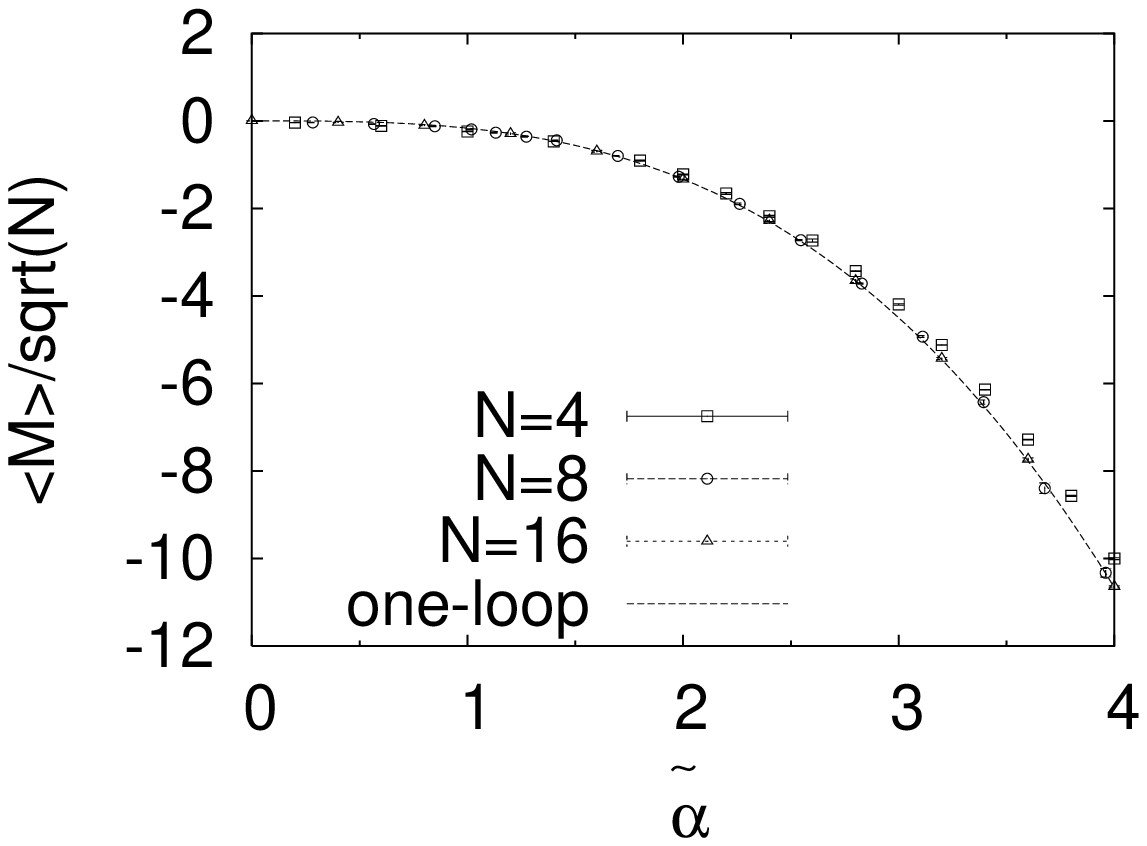,width=7.4cm}
   \caption{Various observables are plotted against ${\tilde \alpha}
   = \alpha \, \sqrt{N}$ for $N=4,8,16$.
   The dashed lines represent the one-loop results at large $N$.}
   \label{observables}}

By performing one-loop calculation around the fuzzy sphere 
$A^{(\stwo)}_{\mu}$,
we obtain the leading large-$N$ behaviors as
(See appendix \ref{one-loop-analysis} for the details)
 \begin{eqnarray}
  \left\langle \frac{1}{N} \, \tr \, (F_{\mu \nu})^{2} \right\rangle &\simeq& 
   \frac{1}{2} \, {\tilde \alpha}^{4} + 6 \ , \label{f-sq-1loop} \\
  \frac{1}{\sqrt{N}} \left\langle M \right\rangle &\simeq& 
  - \frac{1}{6} \, {\tilde \alpha}^{3} + 0 \ , 
  \label{cs-a-1loop} \\
  \frac{1}{N} \left\langle \frac{1}{N} 
  \tr \, (A_{\mu})^{2}
  \right\rangle
  &\simeq& \frac{1}{4} \, {\tilde \alpha}^{2} + 0 \ , \label{a-sq-1loop}
 \end{eqnarray}
where we have introduced the rescaled parameter
  \begin{eqnarray}
   {\tilde \alpha} = \alpha \, \sqrt{N} \ . 
\label{rescaled-tilde}
  \end{eqnarray}
In the r.h.s.\ of eqs.\ (\ref{f-sq-1loop}) $\sim$ (\ref{a-sq-1loop})
the first term represents the classical result, and the second
term represents the one-loop correction.
%% We will postpone the discussion of the quantity
%% $\left\langle \frac{1}{N} \sum_{\mu=1}^{4} \tr \, A^{2}_{\mu} 
%% \right\rangle$ since it involves some subtle issues.

  \FIGURE{
\epsfig{file=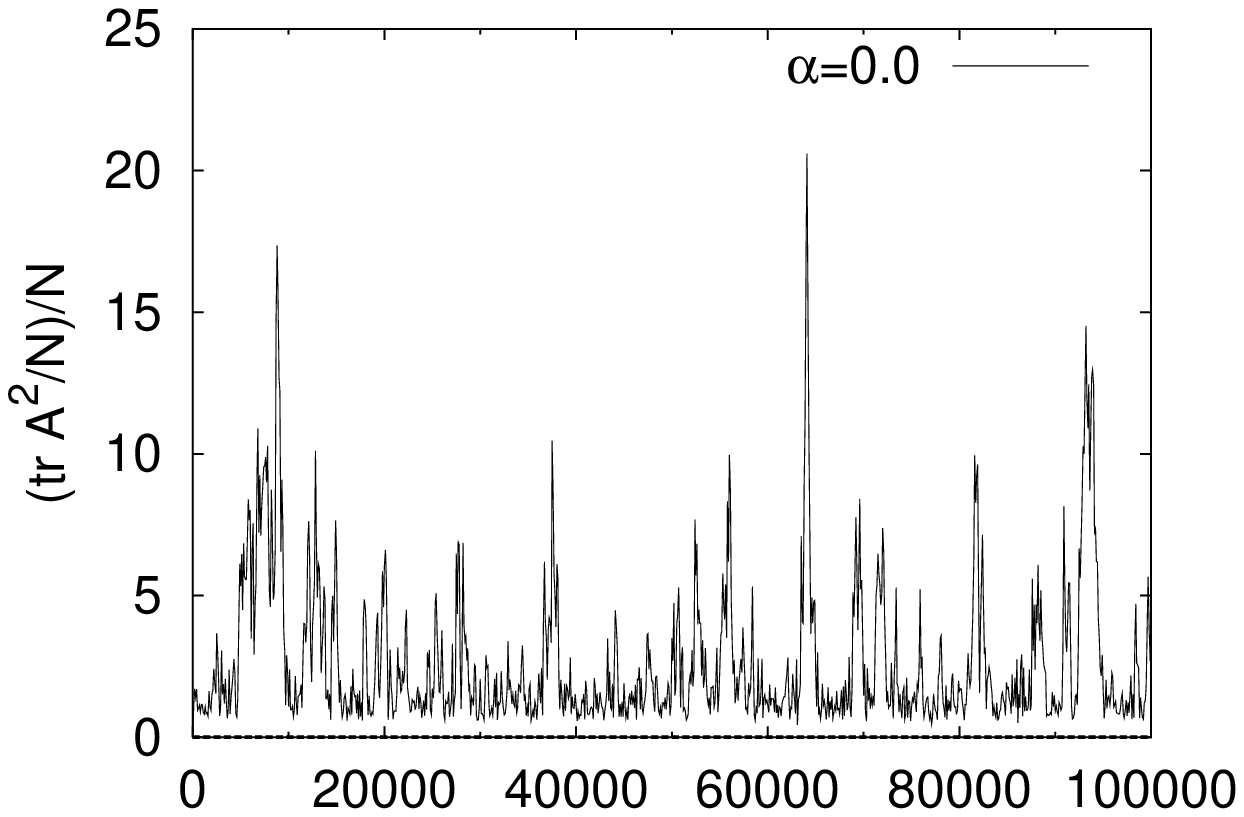,width=7.4cm}
\epsfig{file=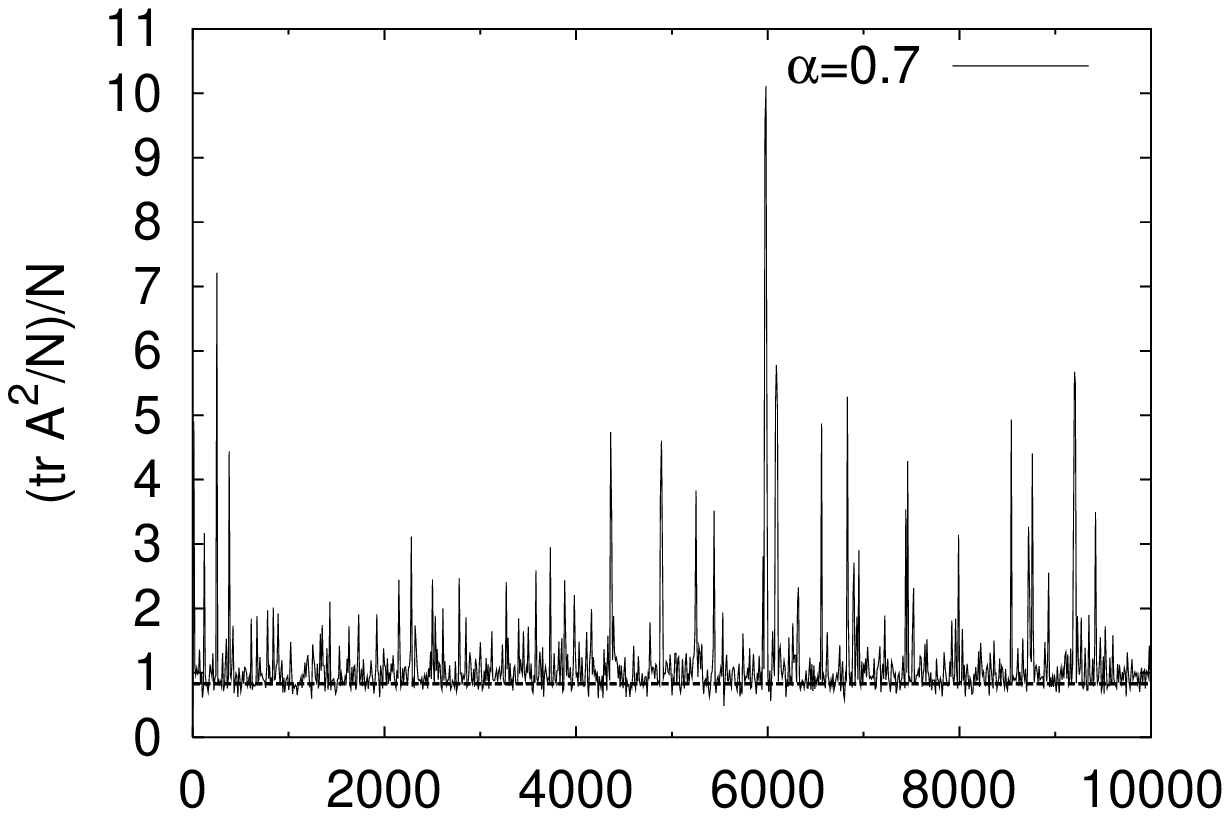,width=7.4cm}
\epsfig{file=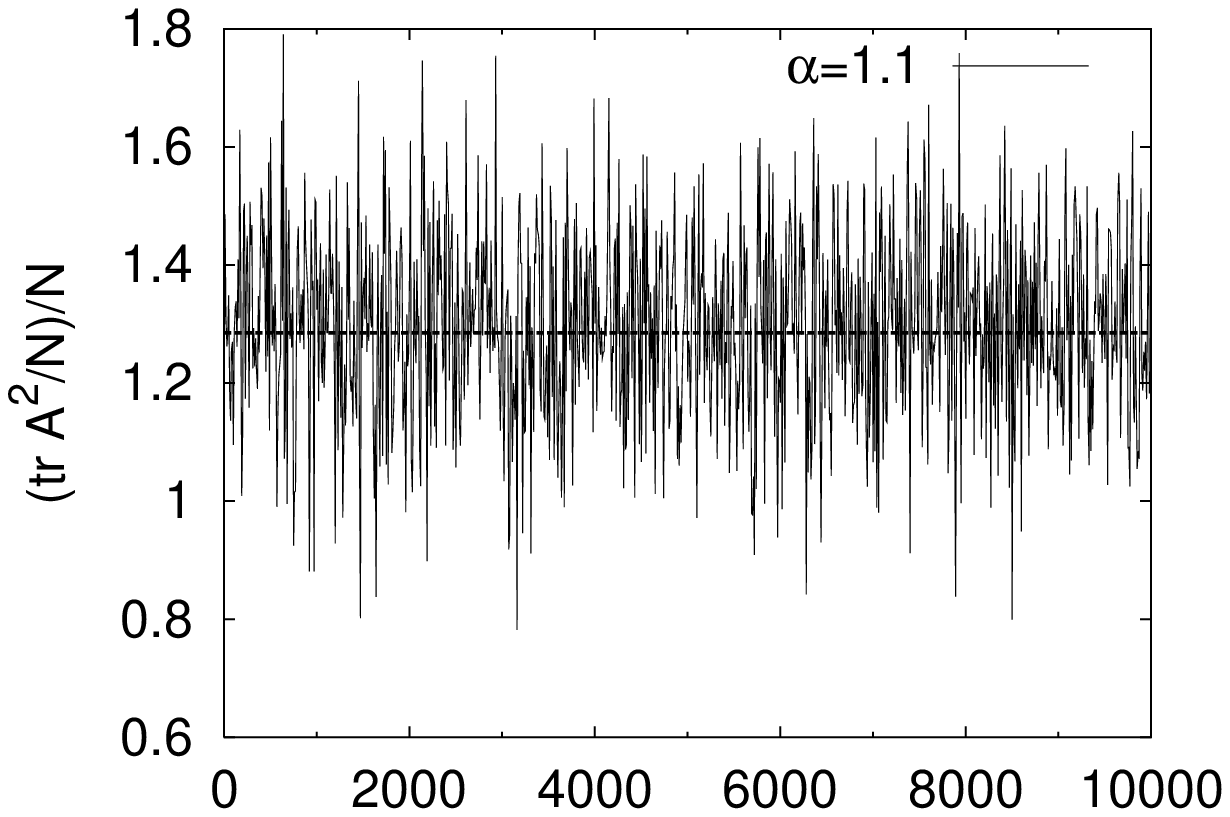,width=7.4cm}
\epsfig{file=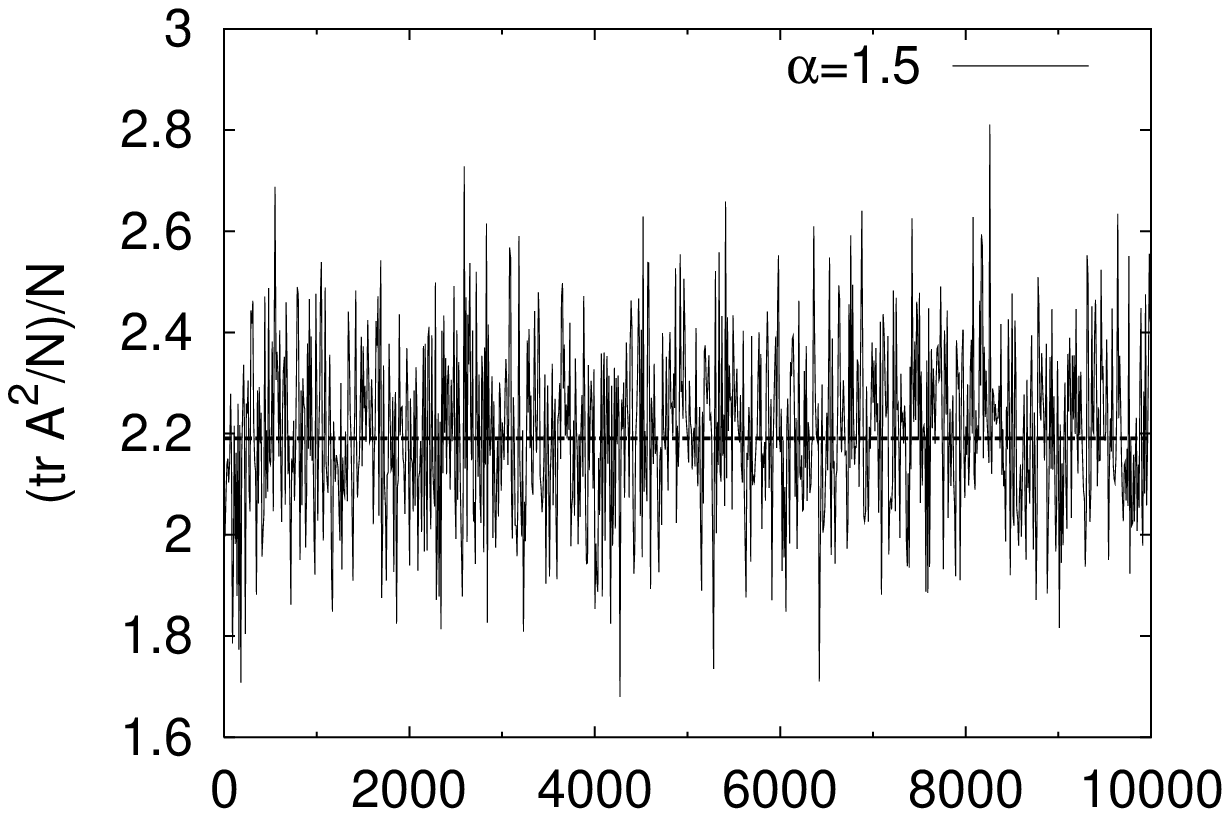,width=7.4cm}

   \caption{The history of $(\frac{1}{N}\, \tr \, (A_\mu)^2)/N$ is shown for
 various $\alpha$ at $N=4$.
 The horizontal lines represent the one-loop results for 
 $\frac{1}{N} \left\langle \frac{1}{N}  \tr \, (A_{\mu})^{2} 
 \right\rangle$.
}
   \label{history_trA2}}

In figure \ref{observables} we plot the results for 
$\left\langle \frac{1}{N} \, \tr \, (F_{\mu \nu})^{2} \right\rangle$ and
$\frac{1}{\sqrt{N}} \left\langle M \right\rangle$
obtained by Monte Carlo simulations.
We find that Monte Carlo data agree with the one-loop results even at 
$\tilde{\alpha}=0$.
This is rather surprising since the expansion parameter in the 
perturbative calculation (at finite $N$) is $\frac{1}{\alpha^4}$.
As we will see shortly, however,
the system actually changes its behavior at
$\tilde \alpha \propto \frac{1}{\sqrt{N}}$,
and the agreement in figure \ref{observables}
below that point should rather be considered as accidental.

Let us then consider the quantity 
$\left\langle \frac{1}{N} \, \tr \, (A_{\mu})^{2} \right\rangle$, 
which we have postponed since
it involves a subtle issue.
At $\alpha=0$ this quantity
%$\frac{1}{N} \left\langle \frac{1}{N} 
%\tr A^{2}_{\mu} \right\rangle$ 
is actually divergent
even for finite $N$, as first observed in numerical studies
\cite{9902113} and further confirmed by ref.\ \cite{Burda:2000mn}.
(Ref.\ \cite{Austing:2001pk} provides some analytical explanation.)
On the other hand, from perturbative calculations around the fuzzy sphere,
we obtain a finite result for finite $N$ 
(See eqs.\ (\ref{tr_A^2}) and (\ref{trA4_ol})).
In order to clarify the situation, let us first look at the history of 
$ \frac{1}{N} \, \tr \, (A_{\mu})^{2}$,
which is plotted in figure \ref{history_trA2} 
for various $\alpha$ at $N=4$.
The horizontal axis represents
the number of ``trajectories'' in the hybrid Monte Carlo algorithm
(See appendix \ref{algorithm}.)
with the parameters $\nu=100$ and $\Delta \tau = 0.01$.
For $\alpha=0$ the history has a lot of spikes,
and these spikes are responsible for the divergence of
$\left\langle \frac{1}{N} \, \tr \, (A_{\mu})^{2} \right\rangle$.
As we increase $\alpha$ the spikes become less frequent, and
their height gets lowered.
At $\alpha \gtrsim 1.1$ the history looks quite regular.
We also looked at the history 
at larger $N$, and find that it becomes regular (no spikes)
%for $\alpha \gtrsim 1.0$ at $N=4$,
for $\alpha \gtrsim 0.5$ at $N=8$ and
for $\alpha \gtrsim 0.3$ at $N=16$.
The transition point $\alpha_{\rm tr}$ is roughly consistent with
$\alpha_{\rm tr} \propto\frac{1}{N}$ 
(i.e., $\tilde{\alpha}_{\rm tr}  \propto\frac{1}{\sqrt{N}}$).
%% We may understand this behavior as follows.
%% Switching on $\alpha$ amounts to reweighting each configuration
%% by the factor $e ^{- S_{\rm CS}}$, which is of order O(1)
%% for configurations with large $ \frac{1}{N} \, \tr \, A^{2}_{\mu}$ 
%% and $e^{{\rm const.}N^2 \tilde{\alpha}^4$ for 
%% fuzzy-sphere like configurations.
%% Therefore the frequency of the spikes is suppressed by 
%% a factor of $e^{-{\rm const.}N^2 \tilde{\alpha}^4$.
%% This means that the transition 

According to ref.\ \cite{0310170}, switching on $\alpha$ does not
change the convergence properties of the matrix integrals.
Therefore, unless there is some special mechanism 
for canceling the leading divergence,
we get $\left\langle \frac{1}{N} \, \tr \,  (A_{\mu})^{2} \right\rangle = \infty$
even for $\alpha \neq 0$.
The reason why we get finite results by 
perturbative calculations should then be that
such a calculation only include the region of the configuration
space near the fuzzy sphere solution so that
it does not take into account the configurations that have
large $ \frac{1}{N} \, \tr \,  (A_{\mu})^{2}$, which may actually
contribute crucially to the vacuum expectation value of that quantity.

  \FIGURE{
    \epsfig{file=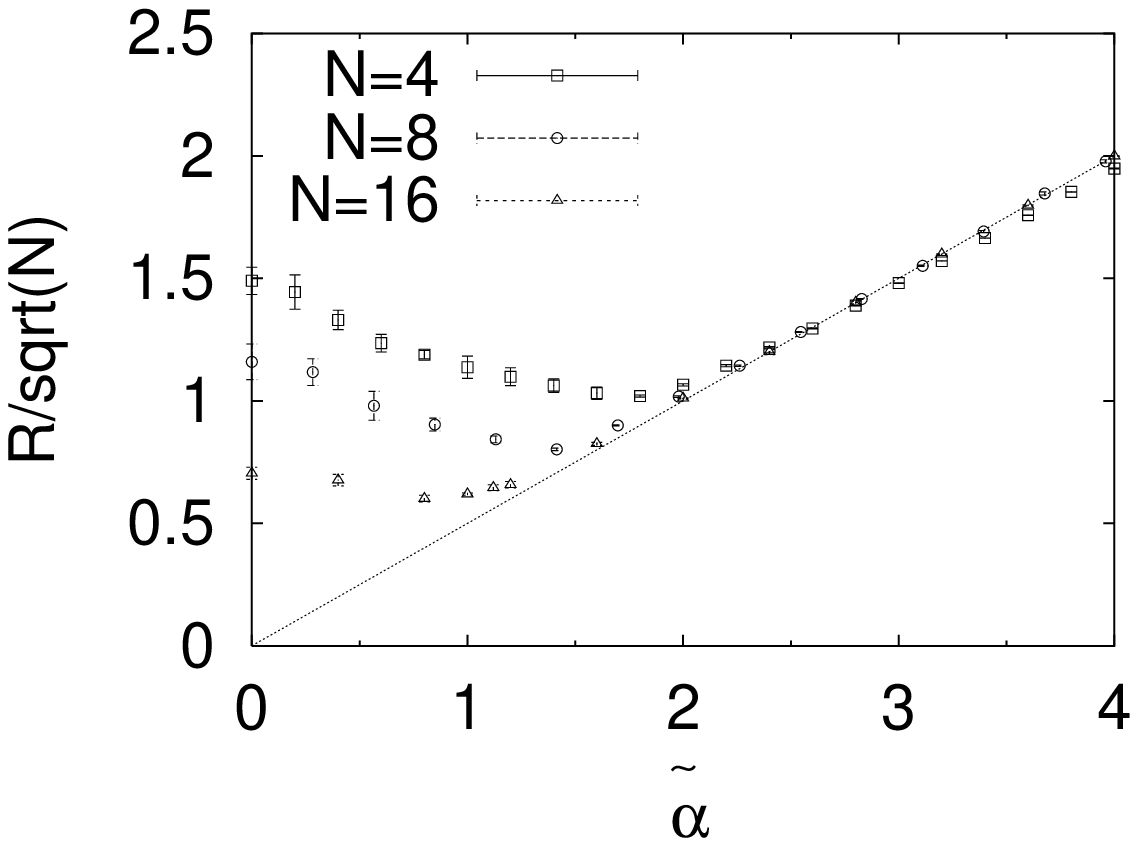,width=7.4cm}
    \epsfig{file=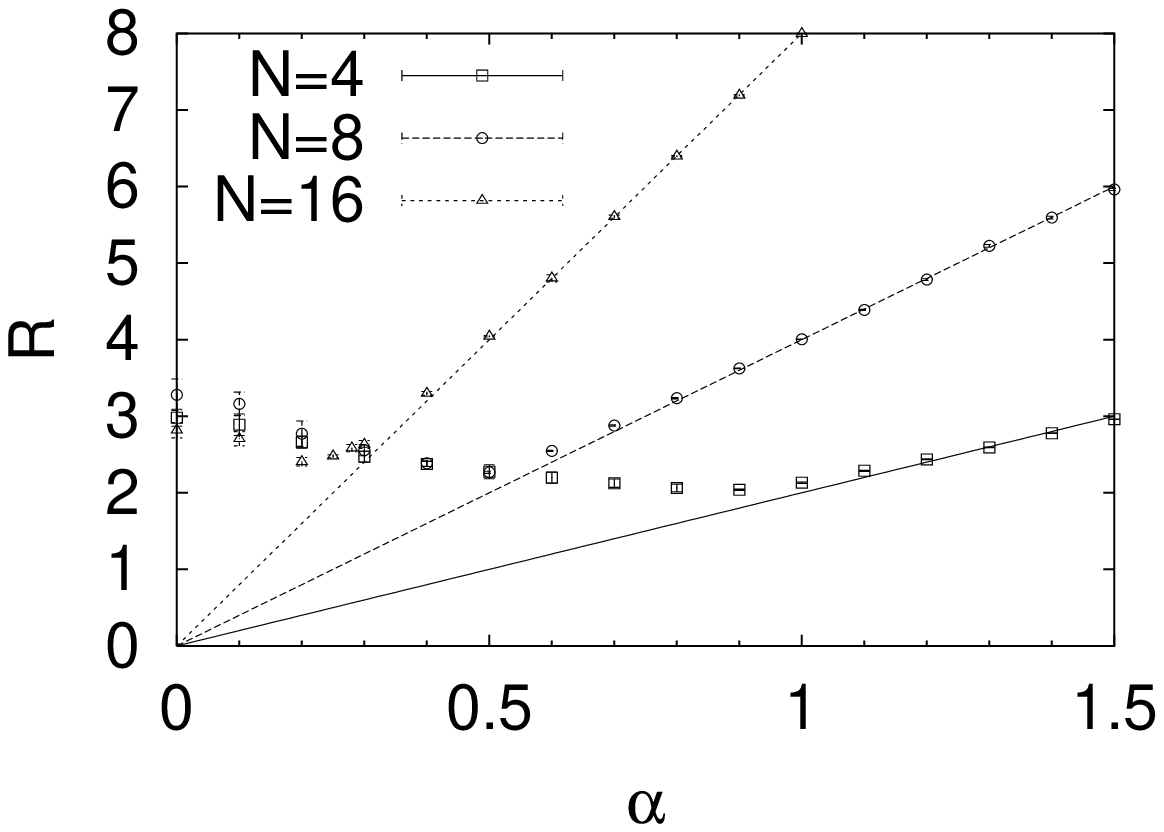,width=7.4cm}
   \caption{Monte Carlo results for 
$\frac{R}{\sqrt{N}}$ are plotted 
against $\tilde{\alpha}$ (left) for $N=4,8,16$.
On the right the same data are plotted in a different scale.
The straight lines represent the classical
result $R=\frac{1}{2} \, \tilde{\alpha} \, \sqrt{N}$ for
the fuzzy sphere solution.
}
   \label{asqsSwithFS}}

%%%%%%%%%%% power-law tail %%%%%

%% In order to clarify the relation of the spikes in the history
%% of $ \frac{1}{N} \, \tr A^{2}_{\mu} $ and the divergence of its
%% vacuum expectation value, 
%% let us consider the 

One can, however, define a finite quantity
\cite{Burda:2000mn}
\begin{eqnarray}
R = \left\langle \sqrt{\frac{1}{N}
\tr \,  (A_{\mu})^{2}} \right\rangle,
\label{refuse}
\end{eqnarray}
which is finite at $\alpha =0$
and behaves as O(1) at large $N$ \cite{Ambjorn:2000bf}
\footnote{Strictly speaking, ref.\ \cite{Ambjorn:2000bf}
studies a slightly different observable,
but the large-$N$ behavior is expected to be qualitatively the same.}
in the present parametrization of the action.
Due to the argument of ref. \cite{0310170}, this quantity is
finite also for $\alpha \neq 0$.
%% However, the divergence is related to the power-law tail of the
%% eigenvalue distribution of the matrices $A_\mu$,
%% and the divergence is actually logarithmic with respect to 
%% a cutoff on the eigenvalues.
%% What one expects to occur in direct Monte Carlo simulations
%% is that the probability of having a configurations with large 
%% $\frac{1}{N} \, \tr A^{2}_{\mu} $ is not suppressed much,
%% so if we can accumulate infinitely large number of configurations,
%% we can eventually see the divergence.
%% However, for finite number of configurations, those rare configurations
%% with large $\frac{1}{N} \, \tr A^{2}_{\mu} $ do not appear,
%% and we obtain finite results.
We plot the Monte Carlo results for $R/\sqrt{N}$
in figure \ref{asqsSwithFS} on the left as a function of $\tilde{\alpha}$
for $N=4$, $8$ and $16$.
At large $\tilde \alpha$ the data agree very well with
the classical result
%\begin{eqnarray}
$R =
% \sqrt{ \left\langle \frac{1}{N}
%\tr A^{2}_{\mu} \right\rangle} \simeq 
\frac{1}{2} \, \tilde{\alpha} \, \sqrt{N} $
% \ .
%\label{hypothesis}
%\end{eqnarray}
for the fuzzy sphere solution.
%% \footnote{
%% In order to calculate $R$ at $\tilde{\alpha} < \tilde{\alpha}_{\rm tr}$
%% we need to accumulate many more configurations than we need
%% at $\tilde{\alpha} > \tilde{\alpha}_{\rm tr}$
%% since the autocorrelation time is much longer due to the spikes.
%% Our present code requires CPU time of the order
%% O($N^6$), and therefore we have so far not been able to obtain 
%% reliable data for $R$ at $N=8,16$ in this regime.}

In figure \ref{asqsSwithFS} on the right we plot
$R$ against $\alpha$.
We expect that $R$ approaches a finite value
in the large-$N$ limit for each $\alpha$ in the small-$\alpha$ regime,
and our data are roughly consistent with this picture.
If we assume the transition to take place at the point where
the fuzzy sphere result $R \simeq \frac{1}{2}\, \tilde{\alpha} \, \sqrt{N}$ 
becomes comparable to the
pure super Yang-Mills behavior $R \simeq {\rm O}(1)$,
the transition point should be 
$\tilde{\alpha} \propto\frac{1}{\sqrt{N}}$, 
which roughly agrees with
the point $\tilde{\alpha}_{\rm tr} $,
where the spikes in figure \ref{history_trA2}
get suppressed.

Thus we conclude that the system actually undergoes
some transition
at $\tilde{\alpha} \propto  \frac{1}{\sqrt{N}}$, below which
the system behaves similarly to 
the pure super Yang-Mills model ($\alpha = 0$).
The quantities in figure \ref{observables} are insensitive to
the transition since their behavior in the pure super Yang-Mills phase
just happens to be the same as in the fuzzy sphere phase.
Note that 
$\langle M \rangle_{\alpha = 0} = 0$ due to parity symmetry
$A_i \mapsto - A_i$, whereas from the perturbative expansion
around the fuzzy sphere, one finds that the one-loop contribution to
$\langle M \rangle$ is absent due to supersymmetry
\footnote{Note that the cubic term, which breaks supersymmetry
softly, does not appear in the relevant one-loop calculation.
This is not the case, however, at higher loop
calculations.},
and as a consequence the result (\ref{o2}) has a smooth extrapolation
to $\alpha =  0$, which agrees with $\langle M \rangle_{\alpha = 0} = 0$.
Since $\langle M \rangle$ and $\langle \frac{1}{N}\, \tr \,  F^2 \rangle$
are related to each other 
through the exact result (\ref{SD2}), the agreement of
the former propagates to that of the latter.
%, as we see explicitly in eq.\ (\ref{exacttrF2_oneloop}).
Thus the success of one-loop results for these quantities
in the small-$\tilde{\alpha}$
regime does not mean that we are still in the fuzzy sphere phase,
but it simply means that these quantities
are insensitive to the transition from the fuzzy sphere phase
to the pure super Yang-Mills phase.

The above results are in striking contrast
to those obtained in the bosonic model \cite{0401038},
where we observed that the fuzzy sphere becomes unstable
at some finite $\tilde{\alpha}$
and various quantities show a hysteresis behavior,
implying a first order phase transition.
%below which the Monte Carlo data disagree with the one-loop results.
%In the present supersymmetric model, we do not see a transition
%at finite $\tilde{\alpha}$.

%%%%%%%%%%

\subsection{Theoretical understanding based on the effective action}

In the previous section we observed that 
the fuzzy sphere is stable in the large-$N$ limit
at any finite $\tilde{\alpha}$ unlike the bosonic model.
Here we would like to provide some theoretical understanding
of the striking difference between the bosonic and supersymmetric
cases based on the one-loop effective action.
For that purpose 
let us consider a one-parameter family of configurations
given by
  \begin{eqnarray}
   \left\{ \begin{array}{ll} A_{i} = \beta \, L^{(N)}_{i} &
  \textrm{~~~for $i=1,2,3$} \ ,  \\
  A_{4} = 0  \ , & ~  
\end{array} \right.
   \label{rescaledFS}
  \end{eqnarray} 
where the fuzzy sphere solution (\ref{fs-solution})
corresponds to $\beta = \alpha$.
The one-loop effective action around (\ref{rescaledFS})
can be calculated along the line described in 
appendix \ref{one-loop-eff-action}, and we get
the result at large $N$ as
  \begin{eqnarray}
 \frac{1}{N^2} \, W_{\rm 1-loop} ^{(\beta)}
  = \left( \frac{1}{8} \, \tilde{\beta}^{4}
  - \frac{1}{6} \, \tilde{\alpha} \, \tilde{\beta}^{3}
  \right) -  \log N \ , \label{one-loop-effective}
  \end{eqnarray}
% (\ref{one-loop-effective}) is derived in detail in appendix 
% \ref{one-loop-analysis}. 
where $\tilde{\beta} = \beta \, \sqrt{N}$.
The one-loop effective action has a minimum
at $\tilde{\beta} = \tilde{\alpha}$ for arbitrary $\tilde{\alpha}$.

In analogous calculations in the bosonic models \cite{0401038,0405277,0506205},
the one-loop contribution gives rise to a term proportional to
$\log \tilde{\beta}$.
Due to this term the (local) minimum disappears below some critical 
$\tilde{\alpha}$, which indeed agrees well with the Monte Carlo results.
In the present supersymmetric case, 
the $\tilde \beta$-dependent one-loop term is absent due to supersymmetry.
Thus we can understand the qualitative difference between the 
bosonic case and the supersymmetric case observed in Monte Carlo 
simulations.

\section{Geometrical structure} \label{a0}

%In the previous section we have seen that the fuzzy sphere
In this section we study the geometrical structure of the 
dominant configurations in the supersymmetric model.
%  We next see more directly that the dominant configurations 
%  have the geometry of the fuzzy $\stwo$.
For that purpose we consider the ``Casimir operator''
  \begin{eqnarray}
   Q = (A_{\mu})^{2}  \label{casimir-q}
  \end{eqnarray}
and define its eigenvalue distribution $f(x)$ as 
  \begin{eqnarray}
   f(x) = \frac{1}{N} \sum_{j=1}^{N} \, 
   \Bigl \langle \delta(x - \lambda_{j})
   \Bigr \rangle \ , \label{f-distribution}
  \end{eqnarray}
where $\lambda_{j}$ $(j=1,2,\cdots,N)$ 
represent the eigenvalues of $Q$.
Let us note 
that $\left\langle \frac{1}{N} \, \tr \,  (A_{\mu})^{2}\right\rangle$
discussed in the previous section
is related to $f(x)$ as
\begin{eqnarray}
 \left\langle \frac{1}{N} \, \tr \, (A_{\mu})^{2} \right\rangle
= \left\langle \frac{1}{N} \, \tr \, Q \right\rangle 
= \left\langle \frac{1}{N} \sum_{j=1}^N \, \lambda_j \right\rangle 
= \int _{0}^{\infty} x \, f(x) \, dx \ .
\label{trA_f_rel}
\end{eqnarray}
In figure \ref{ev-dis} we plot the eigenvalue distribution
for the same set of $\alpha$ and $N$ as in figure \ref{history_trA2}.

\subsection{Power-law tail}

The results for $\alpha = 0$ reproduce the power-law behavior
\beq
f(x) \propto x^{-2} \ ,
\label{power-law-a0}
\eeq
at large $x$, which has been first discovered in ref.\ \cite{9902113}
and studied also in ref.\ \cite{Burda:2005we}.
This is related to the divergence of 
$\left\langle \frac{1}{N} \, \tr \,  (A_{\mu})^{2}\right\rangle$
discussed in section \ref{section:MCresults}.
In fact this quantity is known to diverge {\em logarithmically}
\cite{9902113}, which explains the power in (\ref{power-law-a0})
due to (\ref{trA_f_rel}).
%% From the results in ref.\ \cite{9902113},
%% we find that the divergence is actually logarithmic 
%% in the sense that if we
%% introduce a cut-off by adding a term like 
%% $\frac{1}{\Lambda} \tr (A_{\mu})^{2}$ 
%% in the action, the expectation value
%% $\left\langle \frac{1}{N} \, \tr (A_{\mu})^{2}\right\rangle_{\Lambda}$ 
%% will be O($\log \Lambda$).
%% In order for this to happen,

At $\alpha = 0.7$ the magnitude of the power-law tail becomes
much weaker, but our data are consistent 
with the existence of the power-law tail.
At $\alpha \gtrsim 1.1$
the power-law tail becomes hardly visible,
which corresponds to the disappearance of 
the spikes in the history of $\frac{1}{N}\, \tr \,  (A_\mu)^2$
seen in Fig.\ \ref{history_trA2}.
If we assume that 
$\left\langle \frac{1}{N} \, \tr \,  (A_{\mu})^{2} \right\rangle = \infty$
for $\alpha \neq 0$, as we argued in the previous section,
the power-law tail should be still there, but simply hidden by
the main contribution coming from the fuzzy-sphere-like configurations
\footnote{Within perturbative expansion around
the fuzzy sphere configuration, we expect to obtain a distribution
which decays faster since the $n$-th moment
$\int dx \, x^n f(x) = \frac{1}{N} \langle \tr \,  \{ (A_\mu)^2 \}^n  \rangle $
can be calculated as a finite quantity to all orders.
}.
Considering that those configurations are enhanced by the 
Boltzmann weight $e ^{{\rm const.} \tilde{\alpha}^4 N^2}$
at large $\tilde{\alpha}$ compared with the configurations
that give the power-law tail,
we expect that the magnitude of the power-law tail decreases
as $e ^{-{\rm const.} \tilde{\alpha}^4 N^2}$.
Therefore the power-law tail is expected 
to disappear completely
if we take the large-$N$ limit at fixed $\tilde\alpha$.

    \FIGURE{
\epsfig{file=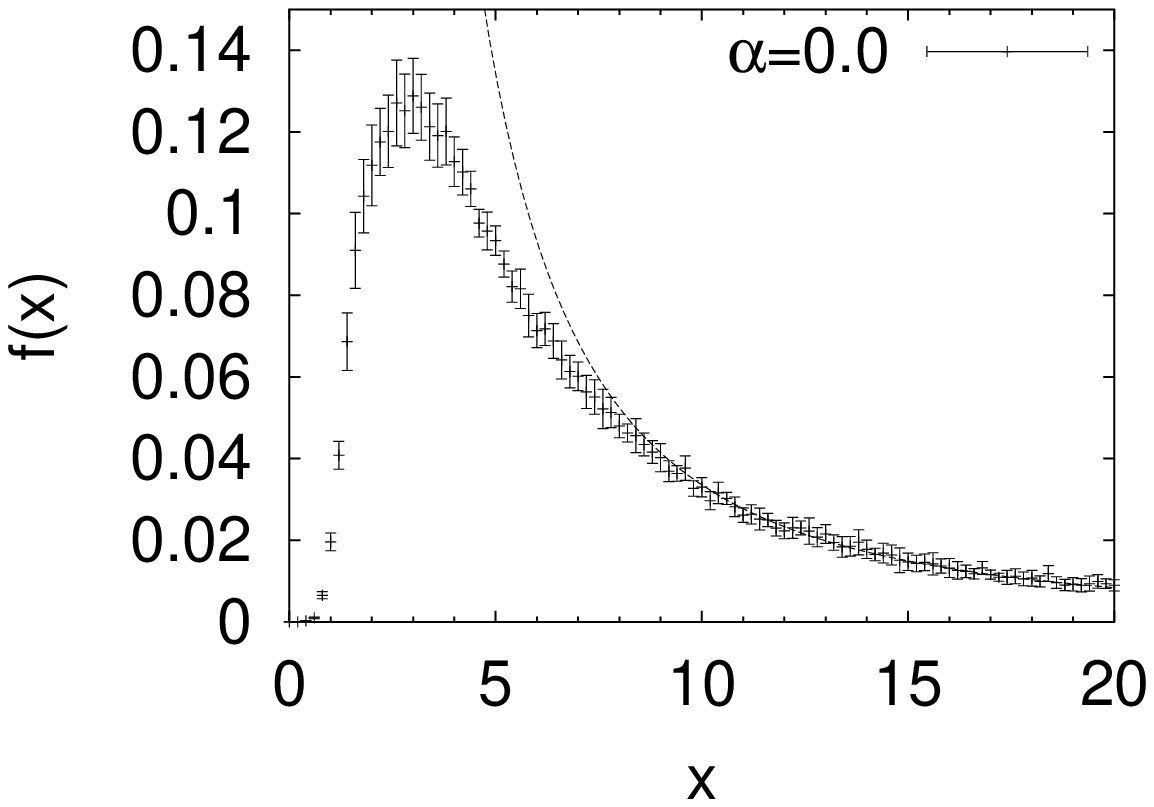,width=7.4cm}
\epsfig{file=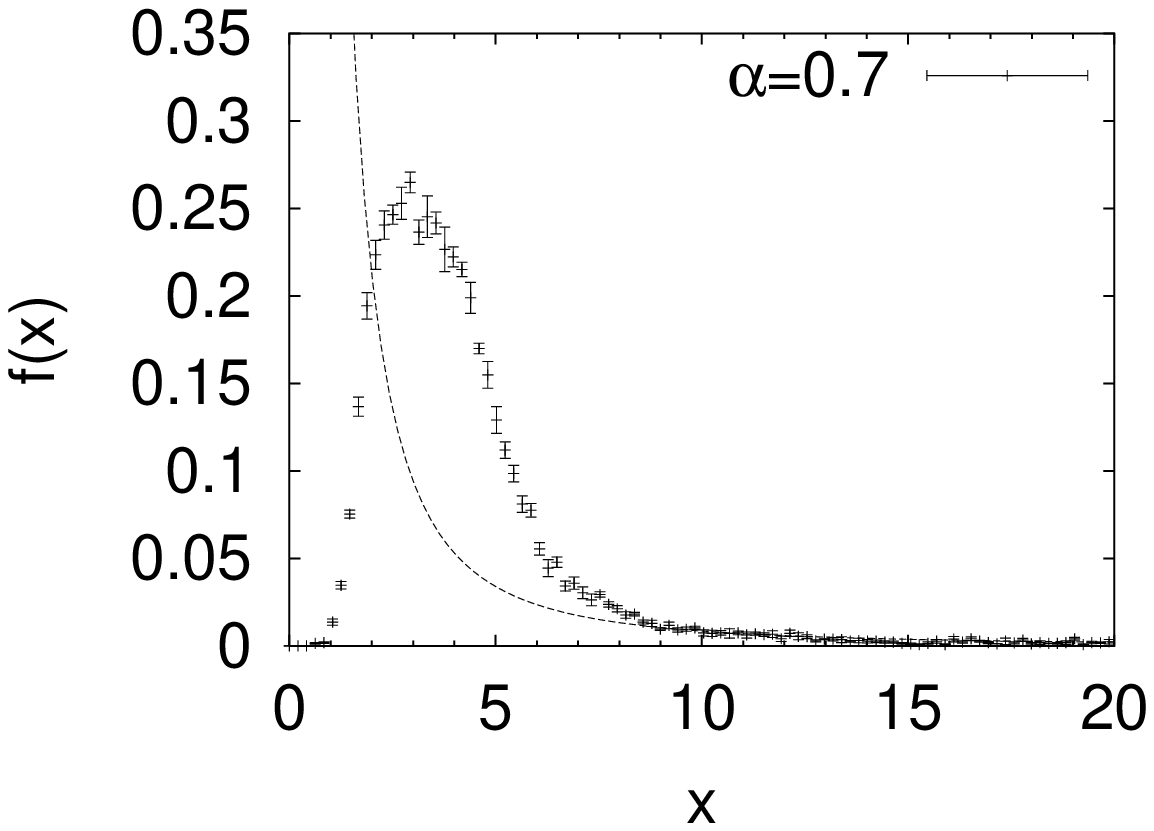,width=7.4cm}
\epsfig{file=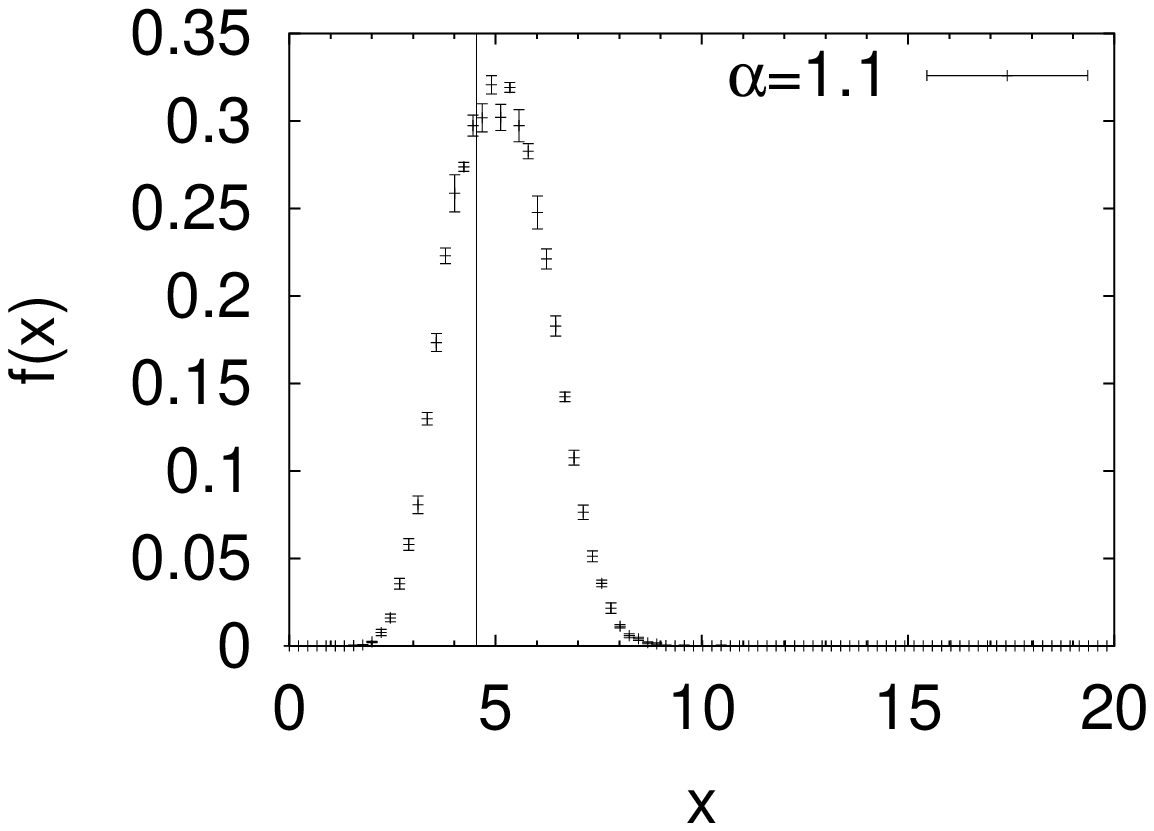,width=7.4cm}
\epsfig{file=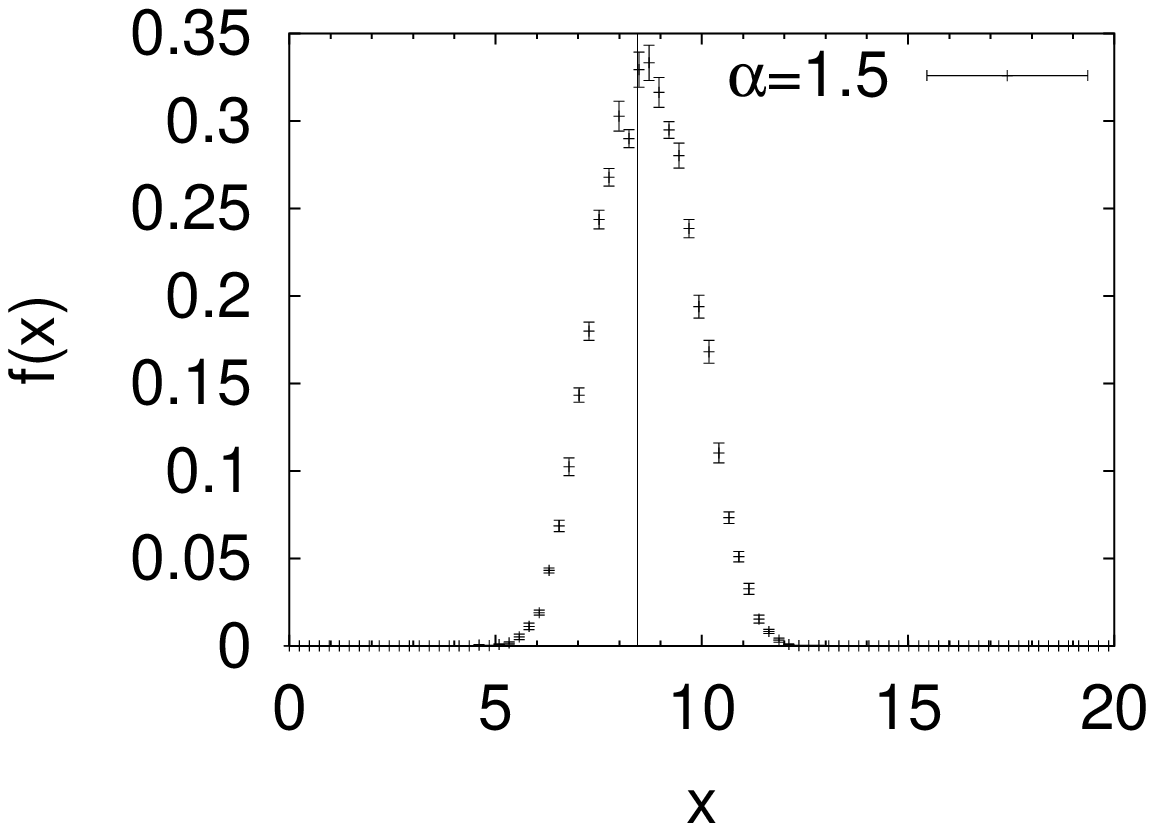,width=7.4cm}
   \caption{The plot of the eigenvalue distribution $f(x)$ of the Casimir 
   operator $Q$ for various $\alpha$ at $N=4$.
   The dashed lines in the two upper figures represent a fit to the
   power-law behavior (\ref{power-law-a0}),
and the vertical lines
in the two lower figures represent the classical
results ($\delta$-function) for the single fuzzy sphere solution.
%the distribution ($\delta$-function).
}
   \label{ev-dis}}

%%%%%%%%%%%%%%%

\subsection{Spherical geometry}

In order to clarify the geometrical structure of the 
fuzzy-sphere-like configurations, 
we decompose the Casimir operator $Q$ as
$ \ Q = Q^{(123)} + Q^{(4)}, \ $ where
\begin{eqnarray}
Q^{(123)}  &=& \sum_{i=1}^3 \, (A_i)^2   \ , \\
Q^{(4)}  &=& (A_4)^2 \ , 
\label{defQ1234}
\end{eqnarray} 
and calculate the eigenvalue distribution for $Q^{(123)}$ and
$Q^{(4)}$, which we denote as $f^{(123)}(x)$ and $f^{(4)}(x)$,
respectively.
Figure \ref{ev-dis2} shows the result for 
$\alpha=1.5$ at $N=4$.
   \FIGURE{
    \epsfig{file=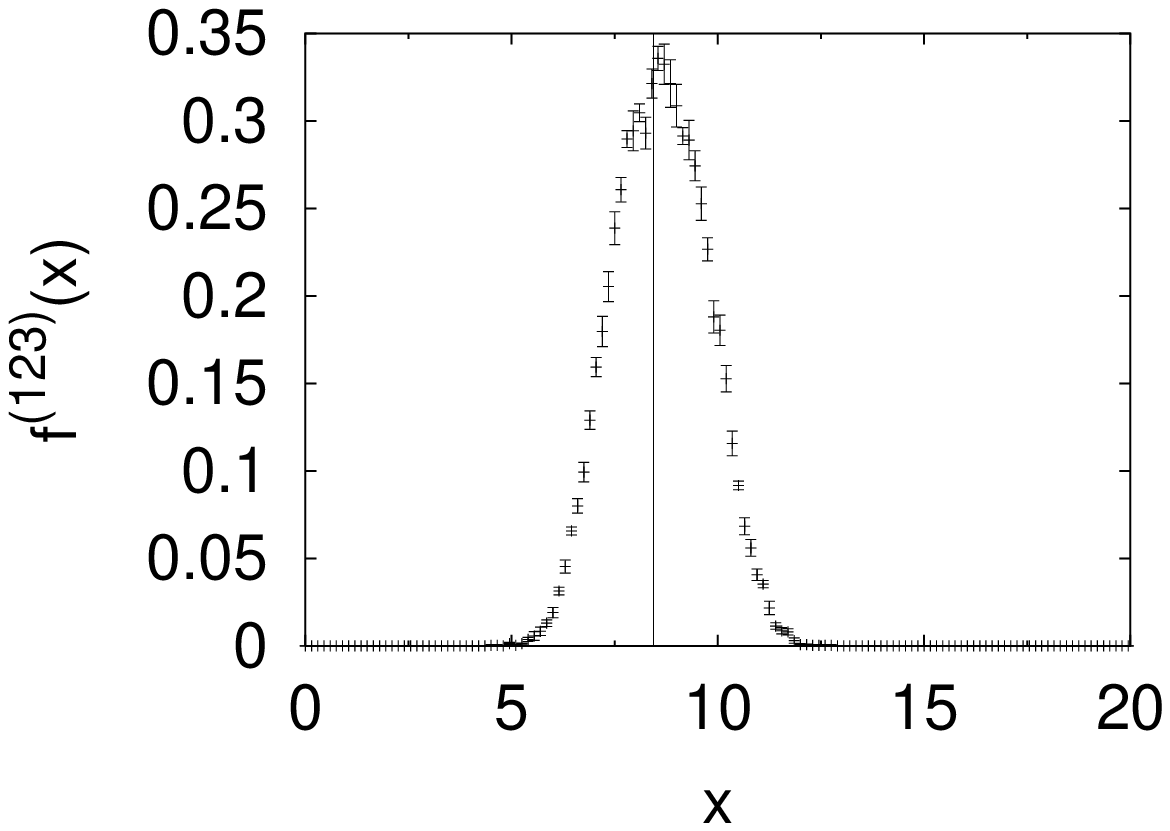,width=7.4cm}
   \epsfig{file=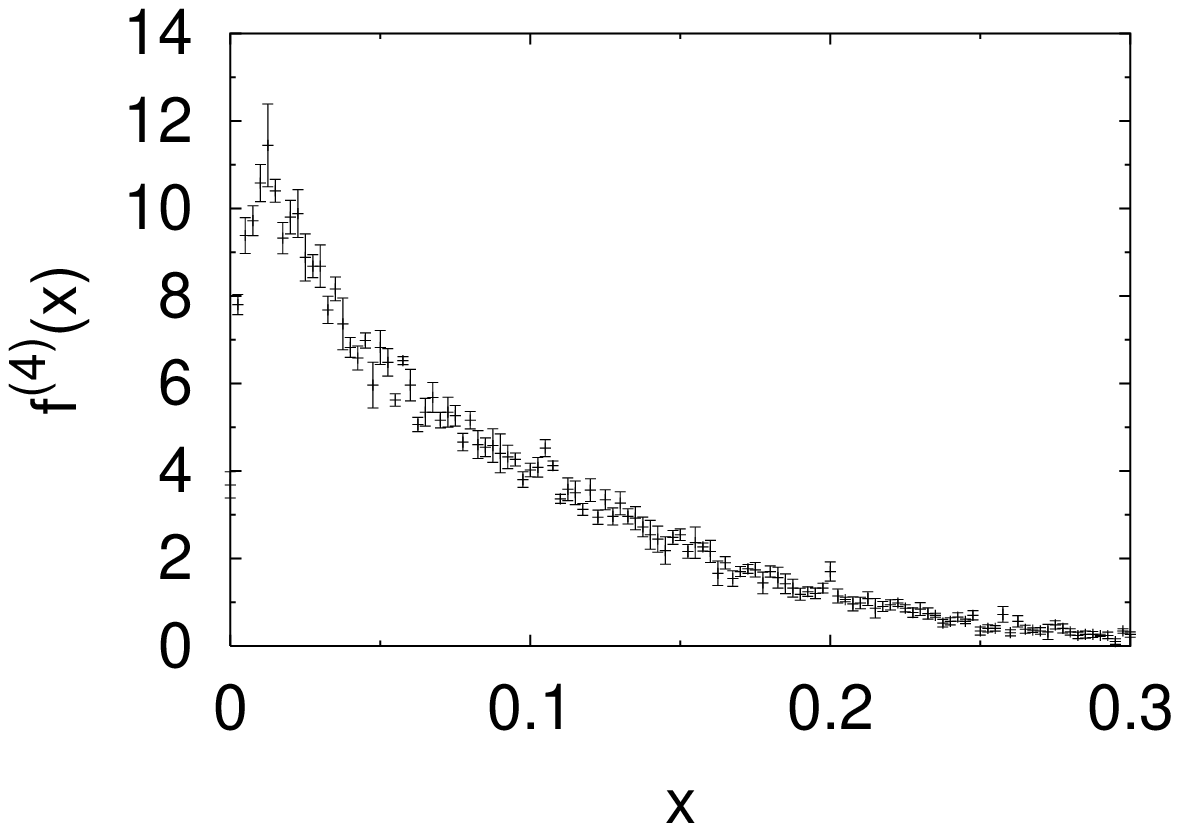,width=7.4cm}
   \caption{The functions $f^{(123)}(x)$ and $f^{(4)}(x)$,
   which are the eigenvalue distributions
   of the operator $Q^{(123)}$
   and $Q^{(4)}$, respectively, are plotted 
   for $\alpha=1.5$ at $N=4$.
   The classical result
   for the single fuzzy sphere solution
%   for $f^{(123)}(x)$ 
   is represented by the
   vertical line ($\delta$-function) in the left figure.
% , whereas the classical result $f^{(4)}(x)$ is given by $\delta (x)$.
}
   \label{ev-dis2}}

The figure on the right
shows that the eigenvalues of $Q^{(4)}$ is quite small,
which can be understood from the one-loop result
(\ref{trA4_ol}) for 
$ \left\langle \frac{1}{N} \, \tr \,  (A_4)^2 \right\rangle
_{\rm 1-loop} $,
which vanishes as O($\frac{1}{N^2}\log N$) in the large-$N$ limit
with fixed $\tilde \alpha$.
%% and compared with the one-loop result (\ref{tr_A^2}) for 
%% $\left\langle \frac{1}{N} \sum_{i=1}^3 \tr \, (A_i)^2 \right\rangle
%% _{\rm 1-loop}$ at large $N$.
As a consequence the distribution $f^{(123)}(x)$ shown
on the left of the figure \ref{ev-dis2} is almost
identical to $f(x)$ shown on the bottom right of figure \ref{ev-dis}.
We also find that $f^{(123)}(x)$
is peaked around the classical result.
Thus we confirm that the dominant configurations 
indeed have the geometry of a 2-sphere.

At $\alpha = 0$ the distribution $f(x)$ has an empty region around
$x=0$. Similar behavior has been observed also in the bosonic
model \cite{0401038}, 
and it can be understood by the uncertainty principle.
Therefore the geometrical structure of the dominant configurations
at $\alpha = 0$ should rather be considered as that of a solid ball.
%% At $\alpha \gtrsim 0.7$
%% the distribution is peaked around the one-loop result for 
%% $\left\langle \frac{1}{N} \sum_{\mu=1}^{4} \tr \,  (A_{\mu})^{2}
%% \right\rangle$.

\section{Dynamical gauge group}
%{Instability of the $k$ coincident fuzzy $\stwo$}
\label{section:multi}

  In fact the equation of motion (\ref{eom}) has a class of
  solutions of the form
  \begin{eqnarray}
 \left\{ \begin{array}{ll}
    A^{(k \stwo)}_{i} = 
  \alpha \, ( L^{(n)}_{i} \otimes {\bf 1}_{k})  & 
  \textrm{~~~for $i=1,2,3$}   \ , \\
  A^{(k \stwo)}_{4}=0 \ , & ~ \end{array} \right. \label{k-fuzzysphere}
  \end{eqnarray}
 where $N = n \, k$.
% and 
% $L_\mu^{(n)}$ represents the $n$-dimensional irreducible representation
% of the SU(2) algebra.
 These solutions represent the $k$ coincident fuzzy $\stwo$, and the
 Casimir operator $Q$ takes the value
  \begin{eqnarray}
   Q = \frac{1}{4} \, (n^{2}-1) \, \alpha^{2} \, {\bf 1}_{N}
  \ , \label{k-casimir}
  \end{eqnarray}
  meaning that the ``radius'' of the fuzzy spheres is given by 
$\rho = \frac{1}{2} \, \alpha \, \sqrt{n^2 - 1 }
\simeq \frac{1}{2 \, k} \, \alpha \, N$,
which becomes smaller as $k$ increases.
By expanding the theory around such a configuration,
  one obtains noncommutative Yang-Mills theory 
  with the U($k$) gauge group \cite{0101102}.

  \FIGURE{
    \epsfig{file=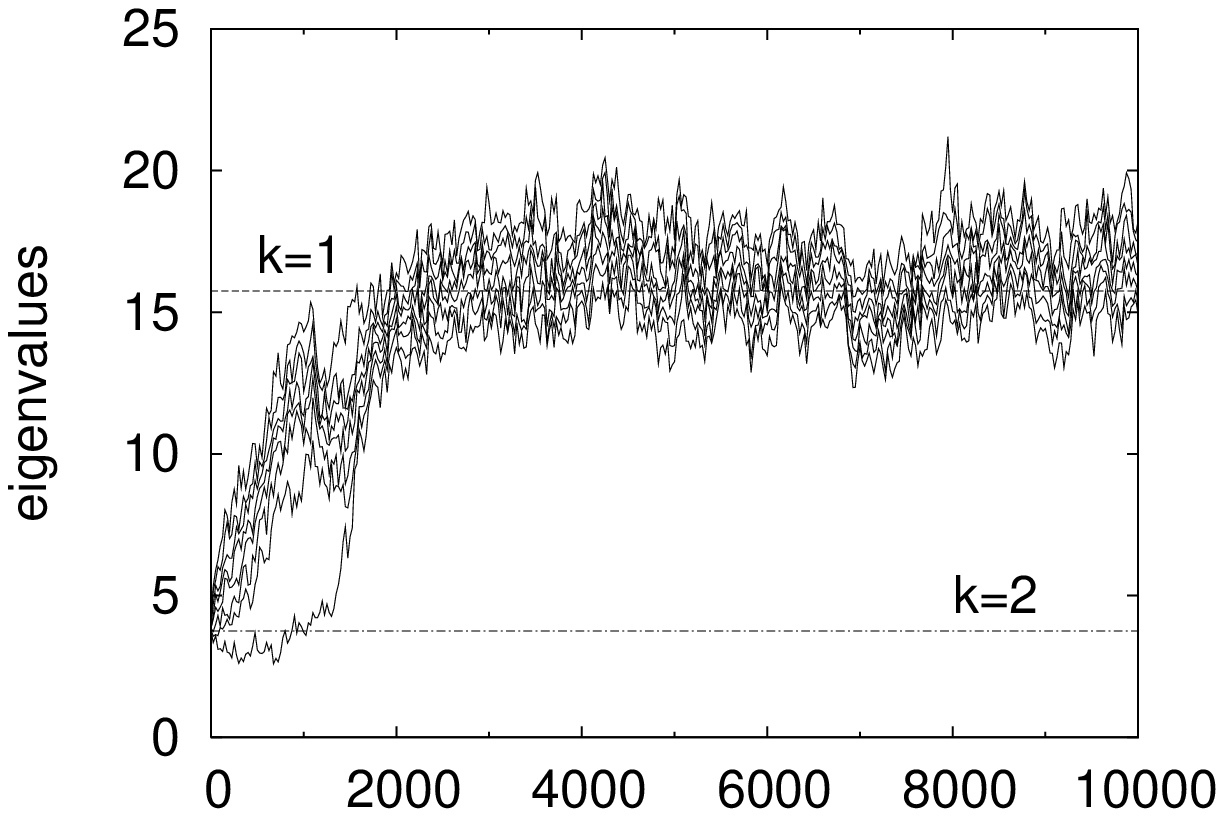,width=7.4cm}
   \caption{The history of the eigenvalues of the Casimir operator $Q$
    obtained for $N=8$, $\alpha=1.0$ using the $k=2$ solution
    as the initial configuration.
    The horizontal lines represent classical results for the single
    fuzzy sphere ($k=1$) and the two coincident fuzzy-spheres ($k=2$), 
   respectively.}
   \label{k2multi-decay}}

  In order to study the stability of such a configuration,
  we perform Monte Carlo simulation for $N=8$ and $\alpha = 1.0$
  using the $k=2$ coincident fuzzy spheres as 
  the initial configuration.
  In figure \ref{k2multi-decay} we plot the history of
  the eigenvalues of $Q$.
 The horizontal axis represents
 the number of ``trajectories'' in the hybrid Monte Carlo algorithm
 (See appendix \ref{algorithm}.)
 with the parameters $\nu=100$ and $\Delta \tau = 0.0001$.
  Thus the $k=2$ multi fuzzy sphere configuration is
  unstable and decays into the single fuzzy sphere.

  This phenomenon can be understood by considering the
  free energy, which is calculated in appendix \ref{one-loop-eff-action}
  up to one-loop. At large $N$ the result reads
  \begin{eqnarray}
   \frac{1}{N^2} \, W_{\rm 1-loop}^{(k)} 
  = - \frac{1}{24 \, k^{2}} \,  {\tilde \alpha}^{4} - \log N \ . 
 \nonumber  
 \label{k-oneloop-effective}
  \end{eqnarray}
  Since the one-loop term represented by the second term 
  is independent of $k$,
  we find that the free energy takes the smallest value for $k=1$.

  Although the conclusion concerning the dynamical gauge group
  is the same as in the bosonic models \cite{0401038,0405277,0506205}, 
  we note that the reasoning is 
  different. In the bosonic models the one-loop term in the
  free energy has the $-\log k^2$ term, which actually favors large $k$.
  However, if one decreases $\tilde{\alpha}$ so that the one-loop term
  becomes more important, the fuzzy sphere solutions disappear
  before the free energy starts to favor $k>1$.
  In the present supersymmetric case, the fuzzy sphere solutions remain
  to be there, but due to the absence of the $-\log k^2$ term, 
  the $k=1$ solution is always favored.

  \section{Conclusion}
\label{section:summary}
 In this paper we have studied the dimensionally reduced 
 4d super Yang-Mills model with an extra Chern-Simons term,
 which incorporates fuzzy spheres as classical solutions.
 We have found that supersymmetry indeed has substantial effects
 on the dynamics of fuzzy spheres.

 While the observables that appear in the action
 change continuously as we vary $\alpha$,
 the model actually possesses
 two distinctive phases, which is demonstrated by a well-defined observable
 $\left\langle \sqrt{\frac{1}{N}\, \tr \,  (A_\mu)^2} \right\rangle$.
 The tail of the eigenvalue distribution changes drastically as one moves
 from one phase to the other.
 In the pure super Yang-Mills phase we observe 
 the same power-law tail as the one known for $\alpha = 0$, but it
 disappears in the fuzzy sphere phase in the large-$N$ limit.
% Accordingly the quantity 
% $\left\langle\frac{1}{N} \sqrt{\tr \,  (A_\mu)^2}\right\rangle$ 
% changes its behavior at some critical point.
 This allows us to identify the critical point quite accurately.
 {}From our Monte Carlo data up to $N=16$ and some theoretical considerations,
 we speculate that the transition point $\tilde{\alpha}_{\rm tr}$ goes to zero
 as $\tilde{\alpha}_{\rm tr} = {\rm O}\left(\frac{1}{\sqrt{N}}\right)$.

 Our results are in sharp contrast to the results obtained for 
 analogous bosonic models \cite{0401038,0405277,0506205},
 where the fuzzy sphere becomes unstable
 at some finite critical $\tilde{\alpha}$.
 A strong first-order phase transition has been observed in the bosonic
 models, and the (lower) critical point obtained by Monte Carlo
 simulation can be reproduced very well from the one-loop effective action.
 In the present paper we have shown that the one-loop term in the 
 effective action vanishes up to an irrelevant constant due to supersymmetry.
 This explains our observation that the fuzzy sphere is stable down to
 vanishingly small $\tilde{\alpha}$ at large $N$.

 We have also studied the dynamical generation of the gauge group.
 In the bosonic case the quantum instability of the fuzzy spheres was
 an obstacle in obtaining a non-trivial gauge group 
 in the true vacuum. In the present supersymmetric case, this instability
 is gone, but we also lost the one-loop term in the free energy
 for the coincident fuzzy spheres, which favors higher multiplicity.
 As a result, we obtain the U$(1)$ gauge group again.
 We note, however, that this conclusion applies only 
 to the fuzzy sphere phase, and whether we can obtain a nontrivial gauge group
 in supersymmetric models {\em without a Chern-Simons term} such as the
 IIB matrix model \cite{9612115} is still an interesting open question.

\acknowledgments
We thank Hajime Aoki, Subrata Bal, Satoshi Iso, 
Kazuyuki Kanaya, Yoshihisa Kitazawa and Dan Tomino
for valuable discussions.
This work was partially 
funded by the 
"Pythagoras" and "Pythagoras II" project,
which is  co-funded by the European Social Fund (75$\%$) 
and Greek National Resources (25$\%$).
The work of T.A.\ and J.N.\ was supported in part by Grant-in-Aid for 
Scientific Research (Nos.\ 03740 and 14740163, respectively)
from the Ministry of Education, Culture, Sports, Science and Technology. 

 \appendix
 \section{Details of the Monte Carlo simulation}
 \label{algorithm}

  In this section we explain the algorithm used for our simulation.
  Our algorithm is similar to the one 
  adopted in ref.\ \cite{Ambjorn:2000bf},
  but the crucial difference is that we make the Metropolis reject/accept
  procedure at the end of each trajectory.
  In the previous algorithm there was a systematic error due to discretization
  required for solving the Hamiltonian equation, and the step size $\Delta\tau$
  for the discretization had to be sent to zero.
  In the present algorithm we do not need such an extrapolation.
  Another difference is that we do not use the noisy estimator
  for estimating the r.h.s.\ of the Hamilton equation since it causes
  some systematic error. Instead we invert 
  the Dirac operator directly using the LU decomposition.
  Each of the two modifications increases the computational effort 
  for making one trajectory (for fixed parameters in the algorithm)
  from O($N^5$) to O($N^6$), which is the price we have to pay
  to make the algorithm ``exact''.

  The ``exact'' algorithm is essentially 
  the hybrid Monte Carlo (HMC) algorithm, 
  which is used in studying the large-$N$ behavior of the phase quenched
  version of the IIB matrix model \cite{Ambjorn:2000dx}.
  In that case the one-loop approximation has been used to decrease
  the computational effort from O($N^6$) to O($N^3$).
  The hybrid algorithms are standard in full QCD simulations, and it 
  is useful also in simulating matrix models
  as demonstrated in refs.\ \cite{Ambjorn:2000bf,Ambjorn:2000dx}.
  If we had used the 
  Metropolis algorithm \cite{Burda:2000mn}
  in the present model, for instance,
  the computational effort would have been O($N^8$).
  Note, however, that simulating matrix models is generally 
  harder than simulating field theories due to the non-local nature
  of the interaction.
  Even in the bosonic case, the computational
  effort is at least O($N^3$), which grows faster than the number of d.o.f.,
  which is O($N^2$). 
%%   We cannot simulate systems with fermionic matrices
%%   with the computational effort of O($N^3$) even with the aid of
%%   the hybrid algorithms, which is in contrast to the situation in QCD, 
%%   where the hybrid algorithms allows us to include dynamical fermions 
%%   without increasing the computational effort by some powers 
%%   of the system size. This is due to the fact that the Dirac operator
%%   in the case of matrix models is non-local

  Let us first recall an explicit form of the fermion determinant
  derived in ref.\ \cite{Ambjorn:2000bf}.
  We define a complete basis for the general complex
  $N \times N$ matrices as
  \begin{eqnarray}
   (t^{a})_{ij} = \delta_{i \, i_{a}} \delta_{j \, j_{a}}  \hspace{3mm}
    (a = 1,2,\cdots,N^{2}) \ ,
  \end{eqnarray}
 where $i_{a},j_{a}$ are integers within the range $1 \leq i_{a}, j_{a} \leq
 N$ satisfying
  \begin{eqnarray}
  a = N \, (i_{a}-1)+j_{a} \ . 
  \label{a-ia-ja}
  \end{eqnarray}
 By taking into account that the fermionic matrices $\psi_\alpha$ and
 $\bar{\psi}_\alpha$ are traceless,
 integration over the fermionic matrices yields
 the fermion determinant $\det {\cal M}$, where 
 the $2 (N^{2}-1) \times 2 (N^{2}-1)$ matrix ${\cal M}$ is given by
  \begin{eqnarray}
   {\cal M}_{a \alpha, b \beta} &=& {\cal M}'_{a \alpha, b \beta}
   - {\cal M}'_{N^{2} \alpha, b \beta} \, \delta_{i_{a} j_{a}} 
   - {\cal M}'_{a \alpha, N^{2} \beta} \, \delta_{i_{b} j_{b}} \ .
  \label{calm-definition}
  \end{eqnarray}
 Here the $2 N^{2} \times 2 N^{2}$ matrix ${\cal M}'$ is defined as
  \begin{eqnarray}
   {\cal M}'_{a \alpha, b \beta} &=& 
   (\Gamma_{\mu})_{\alpha \beta} \, \tr \, ( t^{a} [A_{\mu}, t^{b}]) \ . 
  \end{eqnarray}
 The effective action for $A_\mu$ can be written as
 \begin{eqnarray}
  S_{\rm eff}[A] = S_{\rm b}[A] - \log \, \Det \, {\cal M}[A] \ . 
\label{effective-action}
 \end{eqnarray}

 Following the idea of the hybrid Monte Carlo algorithm,
 we introduce auxiliary bosonic hermitian matrices $P_{\mu}$ and 
 consider the action
 \begin{eqnarray}
   S_{\rm HMC} [P,A] = \frac{1}{2} \, \tr \, (P_{\mu})^{2} + S_{\rm eff}[A] \ .
 \label{HMCaction}
 \end{eqnarray}
 Since $P_{\mu}$ does not couple to $A_\mu$, we retrieve the 
 original model trivially by integrating out $P_{\mu}$.
% original model.
% 
% At first sight the introduction of $P_{\mu}$ seems totally redundant
% since it does not couple to $A_\mu$, but we will see that it enables
% an efficient update of $A_\mu$
% By integrating out the auxiliary variables $P_{\mu}$, we retrieve the 
% original model.
% We simulate the system (\ref{HMCaction}) using the following procedure.
 We regard the action $S_{\rm HMC} [P,A]$ as 
 the Hamiltonian of a classical system described by $A_\mu (\tau)$ and
 its conjugate momentum $P_\mu (\tau)$, where $\tau$ denotes the fictitious
 time of the classical system. Then as an update procedure, we
 may take the old configuration $(P,A)$ as the initial configuration
 $(P(0),A(0))$ and solve the Hamiltonian equation
  \begin{eqnarray}
   \frac{d(A_{\mu})_{ij}}{d \tau} &=& 
   \frac{\partial S_{\rm HMC} }{\partial (P_{\mu})_{ij}}
   = (P_{\mu})_{ji} \ , \label{hamil-x} \\
   \frac{d(P_{\mu})_{ij}}{d \tau} 
   &=& - \frac{\partial S_{\rm HMC} }{\partial (A_{\mu})_{ij}}
   = - \frac{\partial S_{\rm eff} }{\partial (A_{\mu})_{ij}} 
 \nonumber \\
%   = - \frac{\partial S_{b}}{\partial A_{\mu,ij}} - \frac{\partial}{\partial 
%   A_{\mu,ij}} \log \det {\cal M} \nonumber \\
   &=& N \, \Bigl( - [A_{\nu},[A_{\mu}, A_{\nu}]] 
  + 2 \, i \, \alpha \, \epsilon_{\mu \nu \rho}
         A_{\nu} A_{\rho}\Bigr)_{ji} 
   - \Tr \left( {\cal M}^{-1} \frac{\partial 
        {\cal M}}{\partial (A_{\mu})_{ij}} \right)  
  \label{hamil-a}
  \end{eqnarray}
for a finite fictitious time $T$ (this defines ``one trajectory'')
to obtain $(P(T),A(T))$.
The symbol $\Tr$ in (\ref{hamil-a}) denotes a trace
over the $2 \, (N^{2}-1)$-dimensional index, and the derivative
$\frac{\partial {\cal M}}{\partial  (A_{\mu})_{ij}}$ is given explicitly by
 \begin{eqnarray}
   \frac{\partial {\cal M}_{a \alpha, b \beta}}{\partial (A_{\mu})_{ij}}
 = - (\Gamma_{\mu})_{\alpha \beta} \, \Bigl( [t^{b}, t^{a}] \Bigr)_{ji} \ .
 \end{eqnarray}
Since the trace of $A_\mu$ is not conserved during the evolution,
we subtract the trace part 
$ A_\mu'= A_\mu (T) - \left\{ \frac{1}{N} \, 
\tr \,  A_\mu (T) \right\} {\bf 1}$.
%% and flip the sign of the ``momentum''\footnote{This procedure
%% can be omitted in actual simulations since the updated momentum
%% $P_\mu '$ is never used in the subsequent simulation.
%% Here we include it merely for the sake of explanation on reversibility.}
%% $ P_\mu ' = - P_\mu(T)$.
Thus we obtain the updated configuration $(P',A')$,
where $ P_\mu ' = P_\mu(T)$.
Due to the Hamiltonian conservation, this update procedure preserves the 
action $S_{\rm HMC} [P,A]$.
Using also the fact that the transition between $(P,A)$ and $(-P',A')$ 
is reversible, one can readily verify the detailed balance.
After each trajectory, we update the momentum $P_{\mu}$ 
fixing $A_\mu$, which
can be done by simply generating gaussian variables 
since the $P_{\mu}$-dependent part of the action (\ref{HMCaction}) 
is gaussian.
This procedure is necessary to avoid the ergodicity problem.

In actual calculations we have to discretize the Hamiltonian equation
(\ref{hamil-a}). The reversibility of the time evolution can be
preserved by using the so-called leap-frog discretization, but
the Hamiltonian conservation is inevitably violated.
However, we may accept the configuration $(P',A')$ as the updated 
configuration with the probability $\max (1 , e^{-\Delta S_{\rm HMC}})$, 
where $\Delta S_{\rm HMC}  = S_{\rm HMC} [P',A'] - S_{\rm HMC} [P,A]$,
and duplicate the old configuration when rejected.
By adding such a Metropolis accept/reject procedure, we can
preserve the detailed balance.
The step size $\Delta \tau$ for the time evolution
should be small enough to
keep the acceptance rate reasonably high.
The discretized Hamiltonian equation is given by
  \begin{eqnarray}
  (P_\mu^{(1/2)})_{ij}  &=& (P_{\mu}^{(0)})_{ij}
   - \frac{\Delta \tau}{2} \, 
  \frac{dS_{\rm eff}}{d (A_\mu)_{ij}}(A_\mu^{(0)}) \ , \nonumber \\
    (A_\mu^{(1)})_{ij}  &=& (A_\mu^{(0)})_{ij} + \Delta \tau \, 
    (P_\mu^{(1/2)})_{ji}  \ ,
   \label{lf-step1}  \\
   (P_\mu^{(n+1/2)})_{ij} 
 &=& (P_\mu^{(n-1/2)})_{ij}
   - \Delta \tau \, \frac{dS_{\rm eff}}{d (A_\mu)_{ij}} (A^{(n)}_\mu )
   \ , \nonumber \\
    (A_\mu^{(n+1)})_{ij} &=&
   (A_\mu^{(n)})_{ij} + \Delta \tau \, 
    (P_\mu^{(n+1/2)})_{ji}  \ ,  \label{lf-step2} \\
  (P_\mu^{(\nu)})_{ij}  &=& (P_\mu^{(\nu - 1/2)})_{ij} 
   - \frac{\Delta \tau}{2}  \, \frac{dS_{\rm eff}}{d (A_\mu)_{ij}}
(A_\mu^{(\nu)})  \ ,
  \label{lf-step3}
  \end{eqnarray}
where $n=1,2, \cdots , \nu-1$ and $T = \nu \, \Delta \tau$,
and we have introduced the short-hand notation
$P_\mu ^{(r)} = P_\mu(r \, \Delta \tau)$ and 
$A_\mu ^{(r)} = A_\mu(r \, \Delta \tau)$.
At each step of the ``Molecular Dynamics'',
we have to calculate the inverse ${\cal M}^{-1}$,
and at the end of each trajectory, we have to calculate
$\det {\cal M}$.
These are the dominant part of the numerical calculation,
and it requires a CPU time of the order of O$(N^6)$.
%%   since $X_{\mu}$ is updated by the gaussian variables without any reference
%%   to the previous configuration, as we have explained in step 1.

The hybrid Monte Carlo algorithm involves two
parameters $T $ and $\Delta \tau$,
which
%These parameters can be chosen arbitrarily as far as detailed balance
%is concerned, but we 
can be optimized in such a way that the computational
effort for obtaining one statistically independent configuration
is minimized.
The optimization can be done in a standard way \cite{Ambjorn:2000dx}.
%The autocorrelation time in units
%of Molecular Dynamics step depends on
%$T = \nu \Delta \tau$, but not $\Delta \tau$ and $\nu$ separately.
%This allows us to perform the optimization in two steps.
First we fix $T$ and optimize $\Delta \tau$ so that the
effective speed of motion in the configuration space, 
which is given by the acceptance rate times $\Delta \tau$, is maximized.
%% \footnote{Here, 
%% the acceptance rate and its errorbar are defined as follows.
%% After the system is thermalized, we assign $r_{i} = 1.0 (r_{i} = 0.0)$ 
%% ($i=1, 2, \cdots, l$) when
%% the new configuration is accepted (rejected) 
%% in the Metropolis procedure, respectively.
%% We define the "acceptance rate" 
%% as the average of the sequence $\{ r_{i} \}$, and
%% we put the errorbar by 
%% applying the jackknife method to the sequence $\{ r_{i} \}$.}
Using the $\Delta \tau$ optimized for each $T$, we minimize 
the autocorrelation time (in units of ``Molecular Dynamics step'')
with respect to $T$.
For instance, at $N=16$ and $\alpha = 0.0$ we obtain the 
optimal values $\Delta \tau\sim 0.006$ and $T\sim 1.0$.
%We use these values for studying other values of $\alpha$  at $N=16$.

\section{One-loop free energy}
\label{one-loop-eff-action}
In this section we formulate the perturbation theory
around fuzzy sphere solutions,
and derive the one-loop free energy.
We decompose $A_\mu$, $\psi$ and $\bar\psi$ into the classical 
background and the fluctuation as 
\begin{eqnarray}
A_\mu &=& X_\mu + {\tilde A}_\mu \ ,\\
\psi & =& \chi + \tilde\psi  \, , \mbox{~~~}
\bar\psi  = \bar\chi + \tilde{\bar\psi}  \ , 
\end{eqnarray}
and obtain the free energy around the classical solutions 
by integrating over $\tilde A_{\mu}$, $\tilde\psi$
and $\tilde{\bar\psi}$ perturbatively.
Here we take the classical solution to be
the $k$ coincident fuzzy spheres
$X_{\mu}= A^{(k \stwo)}_{\mu}$, $\chi=\bar\chi=0$, 
which includes the single fuzzy sphere as a special case
$k=1$.

In order to remove the zero modes associated with the SU($N$) 
invariance, we introduce the gauge fixing term and the
corresponding ghost term 
\begin{eqnarray}
 & & S_{\rm g.f.}  = - \frac{N}{2} \, \tr \, 
[ \, X_\mu, \, A_\mu \, ]^2 \ , \\
 & & S_{\rm ghost} = - N \, 
\tr \, \left( \, [ \, X_\mu, \bar c \, ]\, [ \, A_\mu, c \, ] \, \right) 
 = N \, \tr \, \left( \, {\bar c} \, 
[ \, X_\mu, \, [ \, A_\mu,\, c \, ]\, ] \, 
\right)  \  ,
\end{eqnarray}
where $c$ and ${\bar c}$ are the ghost and anti-ghost fields 
respectively. 

The total action
$ \ S_{\rm total} = S + S_{\rm g.f.} + S_{\rm ghost} \ $
can be written as
\begin{eqnarray}
S_{\rm total} &=& S_{\rm cl} + S_{\rm kin}+S_{\rm int} \  , \\
S_{\rm cl}&=& N \, \tr \, \left(-\frac{1}{4} \, [X_\mu,X_\nu]^2 
+\frac{2}{3}  \, i \, \alpha \sum_{i,j,k=1}^3 \, 
\epsilon_{ijk} \, X_i \, X_j \ X_k \right) \ , \\
%&~& -N\tr\left( 
%\bar{\chi} \sum_{\mu=1}^4 \Gamma_\mu [X_\mu +{\tilde A}_\mu,\chi] 
%\right) ,  
S_{\rm kin} &=& 
 N \, \tr \,  \left(  - [\tilde A_\mu,\tilde A_\nu]
[X_\mu,X_\nu]   
 + i \, \alpha \sum_{i,j,k=1}^3 \, \epsilon_{ijk} \,
[ \, \tilde A_i , \, \tilde A_j \, ] \, X_k \right. \n  \\
&& \left. -\frac{1}{2} \,  [X_\mu,\tilde A_\nu]^2 
 +   \bar c \, [ \, X_\mu, \, [ \, X_\mu, \, c \, ] \, ] 
 - \tilde{\bar\psi} \,  \Gamma_\mu \, 
  [ \,  X_\mu, \, \tilde\psi \, ]  \, \right) \ , \\
S_{\rm int}&=& N \, \tr \,  \left(
-  [\tilde A_\mu,\tilde A_\nu][X_\mu,\tilde A_\nu] 
- \frac{1}{4} \, [\tilde A_\mu,\tilde A_\nu]
[\tilde A_\mu,\tilde A_\nu]
\right. \n \\
&&  \left. +
\frac{2}{3} \, i \, \alpha \sum_{i,j,k=1}^3 \, \epsilon_{ijk} 
\, \tilde A_i \, \tilde A_j \, \tilde A_k
+ \bar c \,  [ \, X_\mu, \, [ \, \tilde A_\mu, \, c \, ] \, ]
 - \tilde{\bar \psi} \, \Gamma_\mu \,  
 [ \, \tilde A_\mu, \, \tilde\psi \, ] \, 
 \right) \ . 
\end{eqnarray}
The linear terms in ${\tilde A}_\mu$ cancel 
since $X_\mu$ satisfies the classical equation of motion.

Noting that 
the background configuration $X_\mu$ includes a factor of $\alpha$,
we can rescale the fluctuations as $\tilde{A}_\mu \mapsto \alpha \, 
\tilde{A}_\mu$,
$c \mapsto \alpha \, c$, $\bar c \mapsto \alpha \, \bar c$, 
$\tilde{\psi} \mapsto \alpha^{\frac{3}{2}} \, \tilde{\psi}$, 
$\tilde{\bar\psi} \mapsto \alpha^{\frac{3}{2}} \, \tilde{\bar\psi}$
so that all the terms in the total action $S_{\rm total}$ become 
proportional to $\alpha^{4}$.
This means that the expansion parameter of the present perturbation
theory is $\frac{1}{\alpha^{4}}$.

The free energy $W$ is defined by
\begin{eqnarray}
\ee ^{-W} &=& \int  \dd \tilde A \, \dd c 
\, \dd \bar c \, \dd \tilde\psi \, \dd \tilde{\bar \psi} \, 
\ee ^{-S_{\rm total}} \ ,
\end{eqnarray}
which can be calculated as a perturbative expansion
$ \ W = \sum_{j=0}^{\infty} \, W_j \ $, 
where $  \ W_j = \mbox{O}(\alpha^{4(1-j)}) \ $.
The classical part is simply given by 
$ \ W_{0} = S_{\rm b}[X] \ $.
In order to evaluate the one-loop term $W_{1}$,
we note that the kinetic terms can be written as
\beq
S_{\rm kin} =
N  \,  \tr \, \left(
\frac{1}{2}  \, 
\tilde A_\nu  ({\cal P}_\lambda)^2  
\tilde A_\nu  
 +  \bar c \,  ({\cal P}_\lambda )^2 c \right)
-N \, \tr \, \left( 
\tilde{\bar\psi} \Gamma_\mu {\cal P}_\mu \tilde\psi 
\right) \ ,
\label{SQ_fs}  
\eeq
where we have introduced the operator ${\cal P}_{\mu}$
which acts on a traceless $N \times N$ matrix $M$ as
%and ${\cal L}_\mu$ as 
\begin{equation}
{\cal P}_{\mu} M \equiv [X_{\mu}, M] \ .
%= \alpha {\cal L}_\mu M  
\end{equation}
Then the one-loop term can be expressed as
\begin{eqnarray}
W_{1} &=& W_{1,{\rm b}}+ W_{1,{\rm f}}  \ , \\
W_{1,{\rm b}} &=& 
{\cal T}r \, \log \left\{ N \,  ({\cal P}_\mu)^2 \right \} \ , 
\label{W_b,1-loop}   \\
W_{1,{\rm f}}  &=& - {\cal T}r' \, \log 
\, ( \, N  \, \Gamma_\mu \, {\cal P}_\mu \,  )
\ , \label{W_f,1-loop}
\end{eqnarray}
where the symbol ${\cal T}r$ denotes
the trace in the $(N^2-1)$-dimensional 
linear space which consists of traceless $N \times N$ matrices,
and ${\cal T}r'$ includes the trace over spinor indices as well.

\subsection{Single fuzzy sphere}

Let us first consider the single fuzzy sphere $X_{\mu}=A^{(\stwo)}_{\mu}$.
The classical part is given by
\begin{equation}
W_0 = -\frac{1}{24} \, N^2 \, \alpha^4  \, (N^2 -1) \ ,
\label{Scl_IR}
\end{equation}
and the one-loop terms can be written as
\begin{eqnarray}
W_{1,{\rm b}} &=& 
{\cal T}r \, \log 
\, ( \, N \, \alpha^2 \, {\cal Q}  \, ) \ ,  \\
W_{1,{\rm f}}  &=& 
- {\cal T}r' \, \log \, ( \, N \, \alpha \, {\cal D} \, )
\ . 
\end{eqnarray}
The operators ${\cal Q}$ and ${\cal D}$ are defined as
\begin{equation}
{\cal Q} = \sum_{i=1}^3 \, ({\cal L}_i)^2 \ ,
\hspace{1cm}
{\cal D} = \sum_{i=1}^3  \, \sigma_i \, {\cal L}_i \ ,
\end{equation}
where ${\cal L}_i$ 
acts on a traceless $N \times N$ matrix $M$ as
$ {\cal L}_i \, M \equiv [L_i^{(N)}, M]$.
In order to evaluate the one-loop terms,
we need to solve the eigenvalue problem of the operators
${\cal Q} $ and ${\cal D}$.

The eigenvectors of the operator ${\cal Q}$ are given
by the ``matrix spherical harmonics'' $Y_{lm}$
($l = 0,1,\cdots , N-1$ and $m= -l ,\cdots ,l$), 
which span a complete basis of 
the space of $N\times N$ matrices and have the properties
analogous to the usual spherical harmonics such as
%\begin{equation}
%Y_{lm}=Y_{lm}^{(N)} 
%\end{equation}
%
\beqa
\frac{1}{N} \, \tr \,  \left( Y_{lm}^\dag  Y_{l'm'}  \right)
&=& \delta_{l \, l'}\delta_{m \, m'} \ ,\\
Y_{lm}^\dag &=& (-1)^m Y_{l,-m} \ .
\label{Yconjg}
\eeqa
The corresponding eigenvalues are given by $l \, (l+1)$; i.e.,
\beq
{\cal Q} \,  Y_{lm} =  l \, (l+1) \, Y_{lm} \ .
\eeq
Thus the one-loop term from the bosonic contribution 
is obtained as 
\beq
W_{1,{\rm b}} = \sum_{l=1}^{N-1} \, 
(2 \, l+1) \log \left[ \, N \, \alpha^2 \, l \, (l+1) \, \right]  \ .
\eeq
Here $l=0$ has been omitted from the sum since the trace
${\cal T}r$ in (\ref{W_b,1-loop})
should be taken in the space of {\em traceless} $N \times N$ matrices.

In order to solve the eigenvalue problem of the operator ${\cal D}$,
we note that
\begin{eqnarray}
{\cal D} 
= \sum_{i=1}^3 \, ({\cal J}_i)^2- {\cal Q}-\frac{3}{4} \ ,
\label{D_rel}
\end{eqnarray}
where we have defined the ``total angular momentum'' operator 
\begin{eqnarray}
 {\cal J}_i =  {\cal L}_i + \frac{\sigma_i}{2}  \ .
\end{eqnarray}
By making a linear combination of eigenvectors of 
${\cal Q}$ with the eigenvalue $l \, (l+1)$,
we can construct the eigenvectors
of both $\sum_{i=1}^3 \, ({\cal J}_i)^2$ and ${\cal J}_3$ with the eigenvalues
$j \, (j+1)$ and $m$, respectively,
%This can be done by using the Clebsch-Gordan coefficients 
where $j$ can be either $j=l+\frac{1}{2}$ ($l=0 , \cdots , N-1$)
or $j=l-\frac{1}{2}$ ($l=1 , \cdots , N-1$),
and $m$ takes half-integer values in the range $|m| \le j$.
Explicitly, the eigenvectors are
given by the ``matrix spinorial-spherical harmonics''
%${\cal Y}_{l+\frac{1}{2},m}$ and ${\cal Y}'_{l-\frac{1}{2},m}$: 
%
\begin{eqnarray}
{\cal Y}_{l+\frac{1}{2},m} &=&
\sqrt{\frac{l+\frac{1}{2}+m}{2 \, l+1}} Y_{l,m-\frac{1}{2}}
\otimes | \! \uparrow \, \rangle
+
\sqrt{\frac{l+\frac{1}{2}-m}{2 \, l+1}} Y_{l,m+\frac{1}{2}}
\otimes |\! \downarrow \, \rangle \ ,  \\
{\cal Y'}_{l-\frac{1}{2},m} &=&
\sqrt{\frac{l+\frac{1}{2}-m}{2 \, l+1}} Y_{l,m-\frac{1}{2}}
\otimes |\! \uparrow \, \rangle
-
\sqrt{\frac{l+\frac{1}{2}+m}{2 \, l+1}} Y_{l,m+\frac{1}{2}}
\otimes | \! \downarrow \, \rangle \ , 
\end{eqnarray}
for the cases $j=l+\frac{1}{2}$ and $j=l-\frac{1}{2}$, respectively.
Here the symbol $| \! \uparrow \, \rangle$ and $|\! \downarrow \, \rangle$ 
denotes the two-dimensional eigenvectors of $\sigma_3$ corresponding to
the eigenvalues $1$ and $-1$, respectively.
{}From eq. (\ref{D_rel}), the ``matrix spinorial-spherical harmonics''
are also eigenvectors of ${\cal D}$ with the eigenvalues
\begin{eqnarray}
{\cal D} &=&j \, (j+1)-l \, (l+1)-\frac{3}{4}\\
&=&\left\{
\begin{array}{ll}
l& \mbox{~~~for} \quad j=l+\frac{1}{2} \\
-(l+1) & \mbox{~~~for} \quad j=l-\frac{1}{2}  \ .
\end{array}
\right.
\end{eqnarray}
Namely we have the relation
\begin{eqnarray}
{\cal D} \, {\cal Y}_{l+\frac{1}{2},m}&=&
l \, {\cal Y}_{l+\frac{1}{2},m} \ ,\\
{\cal D} \,  {\cal Y'}_{l-\frac{1}{2},m}&=&
-(l+1) \,  {\cal Y'}_{l-\frac{1}{2},m}  \ . 
\end{eqnarray}
Thus the one-loop term from the fermionic contribution 
is obtained as 
\begin{eqnarray}
W_{1,{\rm f}}  
%&=& - {\cal T}r' \log (N \alpha {\cal D} ) 
=
- \left[ \sum_{l=1}^{N-1} \, 2 \, (l+1)
%\sum_{m=-(l+\frac{1}{2})}^{l+\frac{1}{2}}
 \, \log \, ( \, N \, \alpha \, l \, )
+
\sum_{l=1}^{N-1} \, 2 \, l
%\sum_{m=-(l-\frac{1}{2})}^{l-\frac{1}{2}}
\, \log \{ \, N \, \alpha \, (l+1) \, \}  \, \right] \ ,
\end{eqnarray}
where $l=0$ has been omitted from the first sum since the trace
${\cal T}r'$ in (\ref{W_f,1-loop})
should be taken in the space of {\em traceless} $N \times N$ matrices
(and over the spinor indices).
Let us rewrite the above expression as
\begin{eqnarray}
W_{1,{\rm f}}  
= - W_{1,{\rm b}} - (N^2 -1 )  \log N +  \log N  \ .
\label{Wf_Wb}
\end{eqnarray}
Thus the fermionic contribution cancel the bosonic contribution 
up to the $\alpha$-independent constant.
%% , and the one-loop term is 
%% obtained as
%% \begin{eqnarray}
%% W_{1}= W_{1,{\rm b}} + W_{1,{\rm f}} \simeq - N^2 \log N
%% \end{eqnarray}
%% at large $N$.
%
{}From (\ref{Scl_IR}) and (\ref{Wf_Wb}) 
the one-loop free energy for the single fuzzy sphere 
is obtained at large $N$ as
% is therefore given by
\begin{equation}
W_{\rm 1-loop} 
\simeq N^2 \, \Bigl(
-\frac{1}{24} \, \tilde{\alpha}^4 -  \log N \Bigr)  \ .
\end{equation}

\subsection{$k$ coincident fuzzy spheres}
\label{coin_effact}

Next we consider the $k$ coincident fuzzy spheres 
$X_{\mu}= A^{(k \stwo)}_{\mu}$.
The classical part of the free energy is given by
\beq
W_0 = - \frac{1}{24} \, 
\tilde{\alpha}^4 \, (n^2 - 1) \ . \label{multi_cl_action}
\eeq

In order to calculate the one-loop term,
we consider the $n \times n$ 
version of the matrix spherical harmonics
$Y_{lm}^{(n)}$, and define
\beq
Y_{lm}^{(a,b)} \equiv 
Y_{lm}^{(n)} \otimes {\bf e}^{(a,b)} \ ,
\eeq
where ${\bf e}^{(a,b)}$ denotes a $k\times k$ matrix 
whose ($a,b$) element is 1 and all the other elements are zero.
The $N \times N$ matrices $Y_{lm}^{(a,b)}$ form
a complete basis of the space of $N \times N$ matrices,
%% and it has the properties
%% \beqa
%% \tr \left( (Y_{lm}^{(a,b)})^\dag  Y_{l'm'}^{(a',b')}  \right)
%% &=& \delta_{ll'}\delta_{mm'}\delta_{aa'}\delta_{bb'} \ ,\\
%% %
%% (Y_{lm}^{(a,b)})^\dag &=& (-1)^{m_1}(-1)^{m_2} Y_{l,-m}^{(b,a)} \ .
%% \eeqa
and they are the eigenvectors of the operator
$({\cal P}_\mu)^2$ for the present background; i.e.,
\beq
({\cal P}_\mu)^2 \,  Y_{lm}^{(a,b)} 
= \alpha^2 \, l \, (l+1) \, Y_{lm}^{(a,b)} \ .
\eeq
Let us note that the operator $({\cal P}_\mu)^2$ has $k^2$ zero modes
corresponding to $l=m=0$ with arbitrary $(a,b)$.
The mode $ \sum_{a=1}^k \, Y_{00}^{(a,a)}$ corresponds to the trace mode,
which should be omitted due to the traceless condition.
Here we omit the other zero modes by hand, and 
simply consider the non-zero modes.
Then the one-loop term $W_{1,{\rm b}}$ from the bosonic contribution
is obtained as
\beq
W_{1,{\rm b}} =k^2 \sum_{l=1}^{n-1} \, (2 \, l+1)
\, \log \left[ \, N \, \alpha^2 \, l \, (l+1) \, \right]  \ .
\label{coin_ol}
\eeq

The calculation of the fermionic one-loop term proceeds in the same
way except that we have to use the matrix spinorial-spherical harmonics
for each of $k^2$ blocks. In this case we have $2 k^2$ zero modes,
and two of them correspond to the trace mode. We omit the other zero modes
by hand and obtain
\begin{eqnarray}
W_{1,{\rm f}} 
= - k^2 \left[ \sum_{l=1}^{n-1} \, 2 \, (l+1)
%\sum_{m=-(l+\frac{1}{2})}^{l+\frac{1}{2}}
\log \, ( \, N \, \alpha \, l \, )
+ \sum_{l=1}^{n-1} \, 2 \, l
%\sum_{m=-(l-\frac{1}{2})}^{l-\frac{1}{2}}
 \, \log \, \{ \, N \, \alpha \, (l+1) \, \}  \, \right] \ .
\end{eqnarray}
The cancellation between the bosonic contribution and the fermionic
one occurs here as well, and 
the one-loop free energy is given by  
\begin{equation}
W_{\rm 1-loop} \simeq N^2 \, \left(
-\frac{1}{24 \, k^2} \, \tilde{\alpha}^4  -  \log N \right) \ .
\end{equation}
%% \footnote{This finite contribution, which was missed in 
%% ref. \cite{0101102}, is pointed out in ref. \cite{0303120}. 
%% This contribution means that the SUSY is softly 
%% broken in the model with the Chern-Simons term.}

\section{One-loop calculation of various observables}
\label{one-loop-analysis}

In this section we apply the perturbation theory discussed in
the previous section to the one-loop calculation of various
observables which are studied by Monte Carlo simulations in this
paper. We take the background to be $k$ coincident fuzzy
spheres $X_{\mu}= A^{(k\stwo)}_{\mu}$,
but the results for the single fuzzy sphere
can be readily obtained by setting $k=1$.  As in
appendix \ref{coin_effact}, we omit the zero modes for $k \ge 2$.

We note that the number of loops in the relevant diagrams can be less
than the order of ${1}/{\alpha^4}$ in the perturbative expansion since
we are expanding the theory around a nontrivial background.  At the
one-loop level, the only nontrivial task is to evaluate the tadpole
$\langle (\tilde A_\mu)_{ij} \rangle$ explicitly, which, however, turns
out to vanish due to supersymmetry.

\subsection{Propagators and the tadpole}

The propagators for $\tilde A_\mu$, the ghosts and 
the fermion fields are given respectively as
\begin{eqnarray}
\left\langle (\tilde A_\mu)_{ij} (\tilde A_\nu)_{kl}
\right\rangle _0
&=& 
\delta_{\mu\nu} \frac{1}{n} \sum_{ab}
\sum_{l=1}^{n-1} \sum_{m=-l}^{l} \, 
\frac{1}
{N \, \alpha^2 \, l \, (l+1) }
\left( Y_{lm}^{(a,b)}  \right)_{ij}
\left( Y_{lm}^{(a,b)\dag} \right)_{kl}  \ , \\
\Bigl \langle(c)_{km} (\bar c)_{pq} \Bigr\rangle_0 &=& 
\frac{1}{n} \sum_{ab} 
\sum_{l=1}^{n-1} \sum_{m=-l}^{l} \, 
\frac{1}
{N  \, \alpha^2 \, l \, (l+1) }
\left( Y_{lm}^{(a,b)}  \right)_{ij}
\left( Y_{lm}^{(a,b)\dag} \right)_{kl} \ , \\
\Bigl \langle(\psi)_{ij} (\bar\psi)_{kl} 
\Bigr\rangle_0 &=& 
- \frac{1}{n}
\sum_{ab}
\sum_{l=0}^{n-1} \sum_{m=-l-\frac{1}{2}}^{l+\frac{1}{2}} \, 
\frac{1}{N \, \alpha \, l}  
\left({\cal Y}^{(a,b)}_{l+\frac{1}{2},m}\right)_{ij}
\left({\cal Y}^{(a,b)\dag}_{l+\frac{1}{2},m}\right)_{kl} \n \\
&&+
\frac{1}{n} \sum_{ab}
\sum_{l=1}^{n-1} \sum_{m=-l+\frac{1}{2}}^{l-\frac{1}{2}} \, 
\frac{1}{N \, \alpha \, (l+1)}
\left({\cal Y'}^{(a,b)}_{l-\frac{1}{2},m}\right)_{ij}
\left({\cal Y'}^{(a,b)\dag}_{l-\frac{1}{2},m}\right)_{kl} \  , 
\end{eqnarray}
where the symbol $\langle \ \cdot \ \rangle_0$ refers to the
expectation value calculated using the kinetic term $S_{\rm kin}$ 
only.
%
%
%%-------------------------------------------------------
%
The tadpole $\langle\tilde A_i \rangle_{\rm 1-loop}$ ($i=1,2,3$) 
at the one-loop level can be calculated as
\begin{eqnarray}
\langle\tilde A_i \rangle_{\rm 1-loop} 
&=& 
 \left\langle N \tilde A_i  \, 
\tr \,  \left([\tilde A_{\nu} ,\tilde A_{\rho} ][X_{\nu} ,\tilde
A_{\rho} ] \right) \right\rangle_0 
-  \left\langle N \tilde A_i  \, \tr \, 
\left( \bar c \, [X_{\nu} ,[\tilde A_{\nu} ,c] ] \right) 
\right\rangle_0 \n \\
&~&  -  \left\langle N \tilde A_i  \, \tr \, 
\left( \tilde{\bar\psi} \Gamma_\nu 
[\tilde A_{\nu} ,\tilde\psi ] \right) 
\right\rangle_0 \ . \label{tadpole} 
\end{eqnarray}
By redoing the calculation in ref.\ \cite{0403242} 
in the present model, we find that the bosonic contribution
and the fermionic contribution cancel each other
even at finite $N$. We also find that 
$\langle\tilde A_4 \rangle =0 $ to all orders in perturbation
theory due to parity invariance $A_4 \mapsto - A_4$.

%---------------------------------------------------------

\subsection{One-loop results for various observables}
Using the propagator and the tadpole obtained in the previous 
section, we can evaluate various observables easily at the one-loop
level. 
For instance the two-point function
$\left\langle \frac{1}{N} \, \tr \, (A_\mu)^2 \right\rangle$
can be evaluated as follows.
Let us decompose it as
\beq
\left\langle \frac{1}{N} \, \tr \, (A_\mu)^2 \right\rangle
=
\left\langle \frac{1}{N} \, \sum_{i=1}^3 \, \tr \, (A_i)^2 \right\rangle
+ \left\langle \frac{1}{N} \, \tr \, (A_4)^2 \right\rangle \ .
\label{dcompA2}
\eeq
Each term on the r.h.s.\ can be calculated as
\begin{eqnarray}
\left\langle \frac{1}{N} \, \sum_{i=1}^3 \, \tr \, (A_i)^2 \right\rangle
_{\rm 1-loop} 
%
%&=&
%\left\langle \frac{1}{N} \, \tr (A_i)^2 \right\rangle
%_{\rm 1-loop} \n \\
%
&=&\frac{1}{N} \, \sum_{i=1}^3 \, \left[ \tr \, (X_i X_i)
+  2 \, \tr \,  \left( X_i \langle\tilde A_i \rangle _{\rm 1-loop} \right)
%+\langle\tilde A_\mu \rangle _{\rm 1-loop}X_\mu)
+\langle\tr \,  (\tilde A_i)^2 \rangle_0 \right] \n \\
&=&
\alpha^2 \left[
\frac{1}{4}(n^2-1) +0 + \frac{3}{\alpha^4  \, n^2}
\sum_{l=1}^{n-1} \, \frac{2 \, l+1}{l \, (l+1)} \right] ,
\label{tr_A^2} \\
 \left\langle \frac{1}{N} \, \tr \,  (A_4)^2 \right\rangle
_{\rm 1-loop} 
&=& \left\langle \frac{1}{N} \, \tr \,  (\tilde A_4)^2 \right\rangle_0 
= \frac{1}{\alpha^2 \, n^2}
\sum_{l=1}^{n-1} \, \frac{2\, l+1}{l \, (l+1)} \ .
\label{trA4_ol}
\end{eqnarray}
At large $N$ with fixed $\tilde\alpha=\alpha \, \sqrt N$, we get
\begin{eqnarray}
\frac{1}{N} \, \left\langle \frac{1}{N} \, \tr \, (A_\mu)^2
 \right\rangle _{\rm 1-loop} 
\simeq
\frac{1}{N} \, \left\langle \frac{1}{N} \, \sum_{i=1}^3 \, \tr \, (A_i)^2
 \right\rangle _{\rm 1-loop} 
\simeq \frac{1}{4\, k^2} \, \tilde\alpha^2 \ .
\end{eqnarray}
The Chern-Simons term $\langle M \rangle$ can be evaluated as 
\begin{eqnarray}
\langle M \rangle _{\rm 1-loop}
&=& \frac{2 \, i}{3 \, N}  \, \epsilon_{ijk} \, 
\left[  \, 
\tr \, (X_i X_j X_k)
+3\, \tr \,  \left( X_i X_j \langle 
\tilde A_k \rangle_{\rm 1-loop} \right) 
\right]  \nonumber  \\
&=& -\frac{1}{6}\, \alpha^3 \, (n^2-1)  \ .
\label{o2}
\end{eqnarray}
At large $N$ with fixed $\tilde \alpha = \alpha \, \sqrt{N}$, we obtain
\beq
\frac{1}{\sqrt{N}} \, \langle M \rangle _{\rm 1-loop}
\simeq
-\frac{1}{6 \, k^2} \, \tilde{\alpha}^3 \ . 
\label{o2_largeN}
\eeq
The observable 
$\langle \frac{1}{N} \, \tr \,  F^2 \rangle$
can be calculated in a similar manner, but 
we can also obtain it from the exact result (\ref{SD2}) 
using (\ref{o2_largeN}) as
\begin{eqnarray}
\label{exacttrF2_oneloop}
\left \langle \frac{1}{N} \, \tr \,  
(F_{\mu\nu})^2 \right \rangle _{\rm 1-loop}
&=& 6\left( 1-\frac{1}{N^{2}} \right)
- 3 \, \alpha \, \langle M \rangle _{\rm 1-loop} \\ 
&\simeq& \frac{1}{2 \, k^2} \, \tilde{\alpha}^4 + 6 \ . 
\label{o1}
\end{eqnarray}

%------------------------------------------------------

\subsection{Alternative derivation}

Since $\tr \,  F^2$ and $M$ are the operators 
that appear in the action $S$, we can obtain their expectation values
easily by using the free energy 
calculated for the $k$ coincident fuzzy sphere
in Appendix \ref{one-loop-eff-action}.
Let us deform the bosonic action as
\begin{eqnarray}
   S_{\rm b} (\beta_{1}, \beta_{2}, \alpha)
= N^2 \, \left[ \,  \frac{1}{4} \, \beta_{1} \, \tr  \, (F_{\mu\nu})^2
 + \beta_2 \, \alpha \, M \,  \right] 
\label{verydefinition2}
\end{eqnarray}
%
%where 
%\begin{equation}
%M = \frac{2i}{3N} \sum_{i,j,k=1}^3  
%   \epsilon_{ijk} \tr(A_{i} A_{j} A_{k}) .  
%\end{equation}
%
with two free parameters $\beta_{1}$, $\beta_{2}$,
and define the corresponding free energy as
\begin{eqnarray}
  \ee ^ {- W(\beta_{1}, \beta_{2}, \alpha)} 
 =  \int \dd A \, \dd \psi \, \dd \bar\psi \, 
\ee ^{-S_{\rm b}(\beta_{1}, \beta_{2}, \alpha)- S_{\rm f} } \ .
\label{oneloopefb12}
\end{eqnarray}
Then $\langle \tr \,  (F_{\mu\nu})^2 \rangle$ and
$\langle M \rangle$ can be obtained by
\beqa
\left\langle \frac{1}{N} \, \tr \,  (F_{\mu \nu})^2 \right\rangle &=& 
\frac{4}{N^{2}} \left. \frac{\partial W}{\partial \beta_{1}} 
\right|_{\beta_{1} = \beta_{2} = 1}  \ ,
\label{one-loopf-sq2}  \\
\langle M \rangle &=& \frac{1}{\alpha \, N^{2}} 
\left. \frac{\partial W}{\partial \beta_{2}} 
 \right|_{\beta_{1} = \beta_{2} = 1} \ .
\label{one-loopcs-a2} 
\end{eqnarray}
By rescaling the integration variables as
$A_{\mu} \mapsto \beta_{1}^{- \frac{1}{4}} \,  A_{\mu}$, 
$\psi \mapsto \beta_1^{\frac{1}{8}} \, \psi$ 
and 
$\bar\psi \mapsto \beta_1^{\frac{1}{8}} \, \bar\psi$, we get
\beq
   W(\beta_{1}, \beta_{2} , \alpha)
=  \frac{3}{2} \, (N^{2}-1) \, \log \beta_{1} 
+W(1,1,\alpha \, \beta_{1}^{-\frac{3}{4}} \, \beta_{2} )   \ .
\eeq
Using the one-loop result
\beq
W(1,1,\alpha)_{\rm 1-loop}
= -\frac{N^2}{24} \, \alpha^4 \, (n^2 -1) 
- k^2 \, (n^2 - 1) \,  \log N +  k^2 \, \log n  \ ,
\eeq
we can reproduce (\ref{o2}) and (\ref{exacttrF2_oneloop}).

%------------------------------------------------------------

\end{document}